\documentclass[acmlarge]{acmart}
\usepackage{CJKutf8}
\usepackage{stfloats}
\usepackage{makecell}
\usepackage{multirow}
\usepackage{graphicx}
\usepackage{subfigure}
\usepackage{tabularx}
\usepackage{arydshln}
\usepackage[final,commandnameprefix=always]{changes}
\usepackage{threeparttable}

\AtBeginDocument{%
  \providecommand\BibTeX{{%
    \normalfont B\kern-0.5em{\scshape i\kern-0.25em b}\kern-0.8em\TeX}}}

\setcopyright{acmlicensed}
\acmJournal{IMWUT}
\acmYear{2024} \acmVolume{8} \acmNumber{3} \acmArticle{108} \acmMonth{9}\acmDOI{10.1145/3678525}

\begin{document}

\title[AngleSizer]{AngleSizer: Enhancing Spatial Scale Perception for the Visually Impaired with an Interactive Smartphone Assistant}

\author{Xiaoqing Jing}
\email{17888826004@163.com}
\orcid{0009-0002-2412-9979}
\affiliation{%
  \institution{Department of Computer Science and Technology, Tsinghua University}
  \city{Beijing}
  \country{China}
}

\author{Chun Yu}
\email{chunyu@tsinghua.edu.cn}
\orcid{0000-0003-2591-7993}
\affiliation{%
  \institution{Department of Computer Science and Technology, Tsinghua University}
  \state{Beijing}
  \country{China}}
\affiliation{%
  \institution{Key Laboratory of Pervasive Computing, Ministry of Education}
  \city{Beijing}
  \country{China}
}

\author{Kun Yue}
\email{yuek20@mails.tsinghua.edu.cn}
\orcid{0009-0009-2965-559X}
\affiliation{%
  \institution{School of Software, Tsinghua University}
  \city{Beijing}
  \country{China}
}

\author{Liangyou Lu}
\email{luly23@mails.tsinghua.edu.cn}
\orcid{0009-0004-5252-513X}
\affiliation{%
  \institution{Department of Computer Science and Technology, Tsinghua University}
  \city{Beijing}
  \country{China}
}

\author{Nan Gao}
\email{nangao@tsinghua.edu.cn}
\orcid{0000-0002-9694-2689}
\affiliation{%
  \institution{Department of Computer Science and Technology, Tsinghua University}
  \city{Beijing}
  \country{China}
}

\author{Weinan Shi}
\authornote{Corresponding author.}
\email{swn@tsinghua.edu.cn}
\orcid{0000-0002-1351-9034}
\affiliation{%
  \institution{Department of Computer Science and Technology, Tsinghua University}
  \state{Beijing}
  \country{China}
}
\affiliation{%
  \institution{Key Laboratory of Pervasive Computing, Ministry of Education}
  \city{Beijing}
  \country{China}
}

\author{Mingshan Zhang}
\email{zms21@tsinghua.org.cn}
\orcid{0009-0000-2196-304X}
\affiliation{%
  \institution{Department of Computer Science and Technology, Tsinghua University}
  \city{Beijing}
  \country{China}
}

\author{Ruolin Wang}
\email{violynne@ucla.edu}
\orcid{0000-0001-9327-3793}
\affiliation{%
  \institution{University of California, Los Angeles}
  \city{Los Angeles}
  \state{CA}
  \country{USA}
}

\author{Yuanchun Shi}
\email{shiyc@tsinghua.edu.cn}
\orcid{0000-0003-2273-6927}
\affiliation{%
  \institution{Department of Computer Science and Technology, Tsinghua University}
  \city{Beijing}
  \country{China}
}
\affiliation{%
  \institution{Qinghai University}
  \city{Xining}
  \state{Qinghai Province}
  \country{China}
}

\renewcommand{\shortauthors}{Jing et al.}

\begin{abstract}

Spatial perception, particularly at small and medium scales, is an essential human sense but poses a significant challenge for the blind and visually impaired (BVI). Traditional learning methods for BVI individuals are often constrained by the limited availability of suitable learning environments and high associated costs. To tackle these barriers, we conducted comprehensive studies to delve into the real-world challenges faced by the BVI community. We have identified several key factors hindering their spatial perception, including the high social cost of seeking assistance, inefficient methods of information intake, cognitive and behavioral disconnects, and a lack of opportunities for hands-on exploration. As a result, we developed AngleSizer, an innovative teaching assistant that leverages smartphone technology. AngleSizer is designed to enable BVI individuals to use natural interaction gestures to try, feel, understand, and learn about sizes and angles effectively. This tool incorporates dual vibration-audio feedback, carefully crafted teaching processes, and specialized learning modules to enhance the learning experience. Extensive user experiments validated its efficacy and applicability with diverse abilities and visual conditions. Ultimately, our research not only expands the understanding of BVI behavioral patterns but also greatly improves their spatial perception capabilities, in a way that is both cost-effective and allows for independent learning. 

\end{abstract}

\ccsdesc[500]{Human-centered computing~Accessibility systems and tools}
\ccsdesc[300]{Human-centered computing~Interaction design}
\ccsdesc[300]{Human-centered computing~Empirical studies in HCI}

\keywords{Accessibility, Spatial Awareness Training, Blind and Visually Impaired Users, Understanding BVI People}


\maketitle

\begin{figure}[!h]
  \centering
  \includegraphics[width=0.99\textwidth]{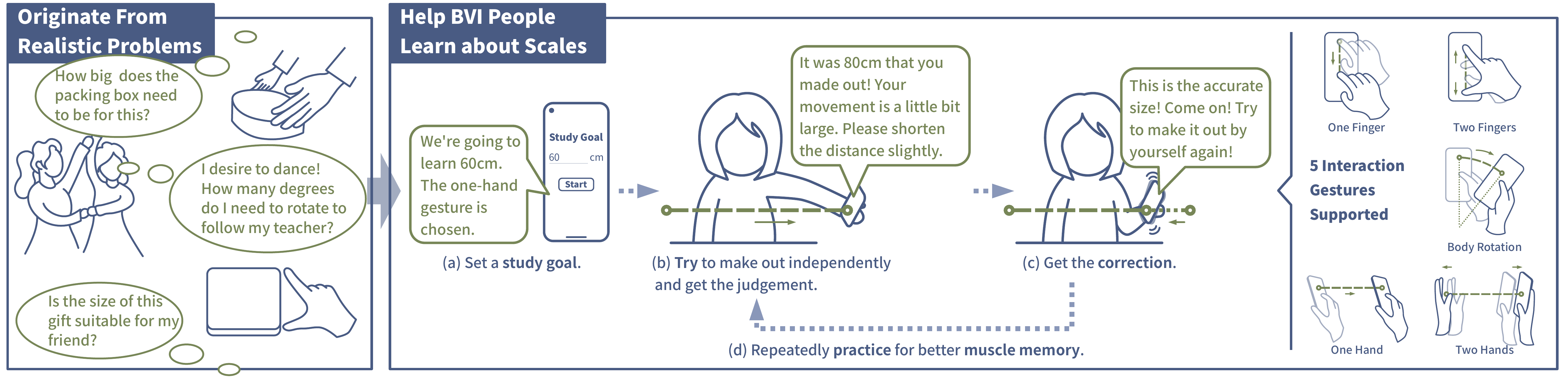}
  \caption{Drawing on insights into real-world challenges faced by the BVI, we proposed AngleSizer. This smartphone assistant supports various interaction gestures, helping BVI users learn about spatial scales through setting study goals, independent exploration, and feedback for corrections. Repeated practice with AngleSizer aids in the BVI developing muscle memory, thereby enhancing spatial perception.} 
  \label{Fig: teaser}
  \Description{Teaser of this paper. From the insight of the real-world challenges faced by BVI, we proposed AngleSizer. It supports various interaction gestures on smartphones, assisting the BVI in understanding scales during the learning process, which includes setting study goals, independent exploration, and receiving corrections. With repeated practice, the blind and low-vision can form muscular memory and enhance their spatial perception.}
\end{figure}

\section{Introduction} 
\label{Cap: Intro}

Understanding scale information, particularly sizes and angles in the human body dimensional range, is a vital aspect of human cognition embedded in our daily lives. From the need to understand item sizes while shopping, to grasping the amplitude of body rotations in dance practice, and even in everyday conversations that involve describing the sizes of objects (see Figure \ref{Fig: teaser}), many scenarios show its importance. This knowledge, forming a key part of spatial perception, is often acquired implicitly by sighted individuals through visual experiences and proprioceptive sources \cite{Rieser1980TheRole}.  However, blind and visually impaired people (BVI) face unique challenges in developing this spatial perception, as they must rely on non-visual senses, requiring additional training to develop this understanding \cite{Jones1975SpatialPI,Rieser1980TheRole}. 

The dominance of vision in the integrated visual-haptic percept, particularly evident in tasks such as judging size, shape, or position \cite{rock1964vision}, underscores the significant impact of visual impairment on BVI individuals' fine motor skills, spatial imagination, and numerical abilities \cite{AssessingVisuallyHandicapped}. This impairment leads to challenges in interacting with their environment and understanding objects. Consequently, BVI individuals often endure negative emotional effects, including low self-esteem and social isolation, which can further compromise their cognitive and emotional well-being \cite{AssessingVisuallyHandicapped, McLinden2020Learn}. Given that approximately 2.2 billion people worldwide are affected by visual impairment \cite{shen2022novel}, addressing these spatial perception challenges is paramount for the holistic development of BVI individuals.

Existing methods for enhancing the BVI's spatial perception show varying degrees of effectiveness. Using a Braille Ruler as a specialized tool for measuring scales in class constitutes direct training. However, its applicability is not universal due to limitations in measurement range dictated by its specifications, and the necessity of interpreting Braille or counting holes to obtain numerical information. Moreover, its portability poses challenges for blind individuals in their daily lives. Developing stable spatial perception abilities and muscle memory necessitates repeated stimulation and practice \cite{Rieser1980TheRole}. However, the Braille Ruler lacks efficient information feedback mechanisms essential for such training. Indeed, digital solutions \cite{TactileCompass,Kassim2016,bastomi2019object} only indirectly guide blind individuals in spatial actions, rather than directly enhancing their cognitive perception of scale.


To address these shortcomings, a deeper understanding of the daily challenges faced by BVI individuals in relation to spatial concepts like size and angles is necessary. This understanding can inform the development of portable, affordable, and targeted tools that enhance the inherent abilities of BVI individuals without creating dependency on the devices. Such tools would ideally bridge the current gaps in training and provide more universally applicable solutions for the BVI community.

In our research, we undertook a comprehensive survey to understand the challenges faced by the BVI community in developing spatial scale perception. Our methodology involved three distinct approaches: distributing questionnaires to 112 BVI individuals, conducting in-depth semi-structured interviews with 12 participants, and performing door-to-door field survey with 6 participants in our local area. Our findings reveal several key barriers in the development of spatial scale perception among BVI individuals, including the high social cost associated with seeking help, the inefficient accumulation of spatial knowledge, disconnects between cognitive understanding and behavioral application, and a general lack of opportunities for exploratory practice. We also gathered insights on the current tactics, behavioral patterns, and specific scale perception needs of BVI users.

Based on the analysis, we designed \textit{AngleSizer}, an innovative assistant for BVI users to acquire and enhance their perception of spatial scales based on size and angle measurement using a smartphone. AngleSizer enables BVI users to perceive and express spatial scales using five natural gestures commonly employed in their daily lives and utilizes Inertial Measurement Unit (IMU) sensors, camera, and touchscreen information to measure spatial dimensions accurately. AngleSizer incorporates three different modules: \textit{Guided Learning}, \textit{Free Exploration}, and \textit{Ability Assessment}, to cater to different scenarios and learning phases. Additionally, AngleSizer offers multimodal feedback channels, including auditory and tactile responses, to support users in exploring spatial dimensions independently. To our knowledge, AngleSizer is the first pioneering digital tool exclusively designed to empower BVI individuals to effortlessly and independently grasp small and medium spatial scales.

To evaluate the teaching effectiveness of AngleSizer, we conducted a 10-day user experiment involving 20 tasks spanning small to medium sizes or angles with 11 participants. Users were encouraged to seamlessly incorporate learning into their daily routines. Their goal was to execute tasks according to AngleSizer's instructions, with a focus on precision, and to leverage the assistant to explore dimensional information in the real-world environment. They spent about 40 minutes a day on this objective. Scale mastery accuracy was evaluated before and after the learning phase. Results indicated AngleSizer's adaptability for diverse visual conditions within the BVI community and favorable teaching outcomes from its natural interactive gestures and instructional design. Structured questionnaire interviews highlighted improved efficiency in information reception, increased curiosity, and enhanced proficiency in discerning intricate scales, fostering self-confidence and communication skills among users. In summary, our contributions are three-folds:

\begin{itemize}
\item  We investigated the specific needs and challenges that BVI individuals face concerning spatial scales and summed up their natural coping mechanisms and strategies used in daily life.

\item  We designed and developed AngleSizer, an innovative smartphone-based teaching assistant. This tool is tailored to assist BVI individuals in learning small and medium spatial scales, by facilitating a convenient, efficient, and cost-effective learning process.

\item We conducted an extensive evaluation of AngleSizer's teaching effectiveness with 11 BVI users. This evaluation includes 20 tasks and daily exploration, thoroughly examining the feasibility and usefulness of AngleSizer in enhancing the understanding of small and medium scales. We also summarised the distinctive experiences and learning characteristics observed during this process. 

\end{itemize}

\section{RELATED WORK} 
\label{Cap: Related Work}

 \subsection{Spatial Space Perception of BVI Individuals}


Spatial scale perception refers to the ability to understand and interpret the relative size between ourselves, objects of interest, and the surrounding environment\cite{10.3389/frvir.2022.672537}. \chadded{Studies report lower task performance of blind compared to sighted people in visuo-spatial imagery \cite{cornoldi1993processing,noordzij2007influence,vecchi1995visuo}, movement memory \cite{dodds1983memory}, and spatial updating \cite{holllns1988spatial,rieser1992visual}.} 
The Japanese scholars' experiment of Bamboo stick length discrimination experiment proved that trained blind children are even better at size discrimination than bright-eyed children, but Angle rotation experiment showed the congenitally blind performed worse on angular discrimination tasks than the blindfolded clear-sighted \cite{1130000798050233216}. Research indicates that while BVI individuals may experience inefficiencies, they do not completely lose their spatial abilities, retaining the potential to develop small and medium-scale perception abilities \cite{Children1, Children3,revesz1950psychology,revesz1950psychology}. Cultivating these abilities relies on senses other than sight, such as touch and hearing, to gauge the spatial attributes of their surroundings \cite{fiehler2009early}.


The majority of BVI individuals typically rely on the following methods to improve spatial perception abilities \cite{Wong2012CanPT}: (1) \textit{Using Body as a Reference}. BVI individuals often use their body as a reference during measurements, enhancing their understanding of spatial scales in relation to their own physical dimensions \cite{ANDREOU2005780}; (2) \textit{Proprioceptive Training}. By utilising auditory \cite{Desprs2005EnhancedSB} and tactile \cite{Shull2015} cues, this method helps BVI individuals in perceiving and navigating spaces without visual input. However, these methods vary in effectiveness and lack a standardized approach. This results in diverse experiences and levels of spatial perception among ordinary BVI people, often leading to inefficiencies in spatial abilities.

Some BVI population has access to specialized training in blind schools. In these settings, students are taught using tools such as 3D teaching aids \cite{3DPrinted, Printing2017, Orly2018, ModelingV, Ottink2022, Enhancing2013,lahav2006using} and braille rulers \cite{brule2021beyond} for tactile exploration. For instance, teachers use graphic tools to introduce the concepts like corners, edges, diagonals, centerlines and curves. The BVI students learn by touching geometric shapes like squares, rectangles and circles, which helps establish correct scale concepts (see \textit{Perkins School for the Blind}\footnote{\textit{Perkins School for the Blind}: \url{ https://www.perkins.org/resource/tips-and-strategies-make-geometry-accessible/}} for an example). However, these exercises rely on the correction of sighted teachers or the reading of braille, and the mechanism of limited learning content, low information acquisition efficiency, and full correction limit the effect of ability cultivation. As a result, major challenges arise once these individuals leave the support of school resources. Due to the precision required in perceiving spatial scale, muscle memory diminishes upon cessation of practice, with an increasing amount forgotten over time \cite{park2013learning,schmidt2018motor,Rieser1980TheRole}.

To the best of our knowledge, there are very few digital teaching tools that address the above challenges. Tools like \textit{Salome} \cite{SALOME} and \textit{Click} \cite{Clicks} provide tactile experiences for feeling geometry, devices like \textit{Compass} \cite{TactileCompass, Kassim2016} and glasses \cite{bastomi2019object} offer vibration feedback and voice prompts to aid in understanding angles and distance for developing spatial perception. These tools, however, often fail to link conceptual knowledge with practical application, a gap that becomes evident once BVI individuals lose access to them. Furthermore, the high cost of these resources and their limited availability pose significant barriers to learning. 

\subsection{Touching and Gestures Expression of BVI Individuals}

Understanding spatial scale through touch is a crucial skill for BVI individuals, yet accurately expressing these scales in daily life presents additional challenges. Gestures play a vital role in communication, as found by Dim et al. \cite{dim2014designing}, who observed that gestures can effectively complement speech, particularly when replicating everyday actions or symbolizing objects \cite{dim2014designing}. For instance, it's common for people to use hand gestures that visually mimic the size or shape of objects during discussions. 

This aspect of communication is also pertinent to BVI individuals. Studies by Kane et al. \cite{Kane2011UsableGestures} have explored how blind people use gestures in interacting with touchscreen devices, demonstrating their adaptability to gestural interfaces \cite{Kane2011UsableGestures}. Further research by Kennedy and Davidson  \cite{kennedy1993drawing,davidson1972haptic} highlights that BVI individuals can discern degrees of curvature and spatial relationships through gestures, such as opening and closing their arms or using their palms and fingers. This suggests that integrating gesture-based learning with tactile training can greatly assist BVI individuals in mastering scale perception, thereby reducing communication barriers related to spatial concepts in everyday life. The potential of gestures as a tool for expression and understanding in the BVI community underscores the importance of incorporating such techniques into their training and education.

\subsection{Recognising Spatial Information with Digital Devices} \label{Cap:2.3 Spatial Rec}

Digital devices employ various methods to recognize and express spatial information, which is crucial for applications aiding the BVI. Simple approaches use the device's dimensions as a frame of reference. Applications like \textit{Ruler}\footnote{\textit{Ruler} by NixGame: \url{https://play.google.com/store/apps/details?id=org.nixgame.ruler}} demonstrate this by enabling users to measure distances using ruler markers displayed on the screen. More advanced methods track the device's orientation and motion, calculating movement, rotation, or position relative to other objects. This often relies on sensors like Infrared Radiation (IR) \cite{LineChaser}, LIDAR \cite{10.1145/3411825, 8429597, 8419727}, and ultrasonic technology \cite{Bai2018, Indriya, 8291597, 7885319, 10.1007/978-3-319-56538-5_36}. However, such methods typically require specialized devices or certain smartphone models with such hardware features \cite{LineChaser, CorridorW}.

Recent advances in Simultaneous Localization and Mapping (SLAM) and Visual Inertial Odometry (VIO) offer a less hardware-intensive solution for spatial recognition \cite{10.1145/3098279.3125437}. These technologies enable spatial recognition using standard smartphone features like cameras and Inertial Measurement Units (IMUs), which reduces the need for additional hardware. Frameworks like \textit{ARCore}\footnote{\textit{ARCore}  by Google: \url{https://developers.google.com/ar}} and \textit{MediaPipe}\footnote{\textit{MediaPipe} by Google: \url{https://developers.google.com/mediapipe}} leverage standard smartphone features for augmented reality and object detection, enhancing spatial perception capabilities in modern smartphones.


Existing tools for enhancing spatial perception in BVI individuals often lack practical application and are limited by accessibility and cost. Our research fills this gap by developing an application that integrates gesture-based learning with advanced smartphone technology. This approach not only ensures a more holistic understanding of spatial concepts for BVI individuals but also provides a cost-effective and widely accessible solution. By focusing on practical application and easy integration into daily life, our work stands to make a significant contribution to improving the independence and quality of life for BVI individuals.


\section{Study 1: Understanding Spatial Perception of BVI Individuals} \label{Cap:Preliminary research}


Spatial perception of BVI individuals is substantially influenced by inherent factors like visual memory. Yet, everyday life experiences play a crucial role in either aiding or hindering the development of these perceptual skills \cite{Lewis2002NewMF}. To delve into this, we conducted a preliminary study focused on understanding the challenges BVI individuals face in spatial perception within real-world contexts.

\subsection{Study Methods} \label{Study methods}


To effectively study the spatial perception of BVI individuals, we employed three research methods: online questionnaires, semi-structured interviews, and door-to-door (DTD) surveys. This combined approach balances the extensive reach of questionnaires with the in-depth insights of interviews and DTD surveys, ensuring both broad representation and detailed analysis in our findings.

\subsubsection{\textbf{Online Questionnaires.}}

Our online questionnaire (see Appendix \ref{app:questionnaire} for details), commenced by gathering demographic data including gender, age, geographic location, visual condition, and educational background. This is crucial to ensure the representativeness of our sample and to set the stage for more personalized, in-depth interviews. Participants then assessed the frequency of scale-related issues and their own scale perception abilities through a series of 5-point Likert scale questions. This was followed by multiple-choice questions exploring strategies for managing scale-related challenges. Finally, open-ended questions allowed for sharing of additional difficulties and strategies in spatial perception. 

\begin{figure}
  \centering
  \includegraphics[width=0.72\textwidth]{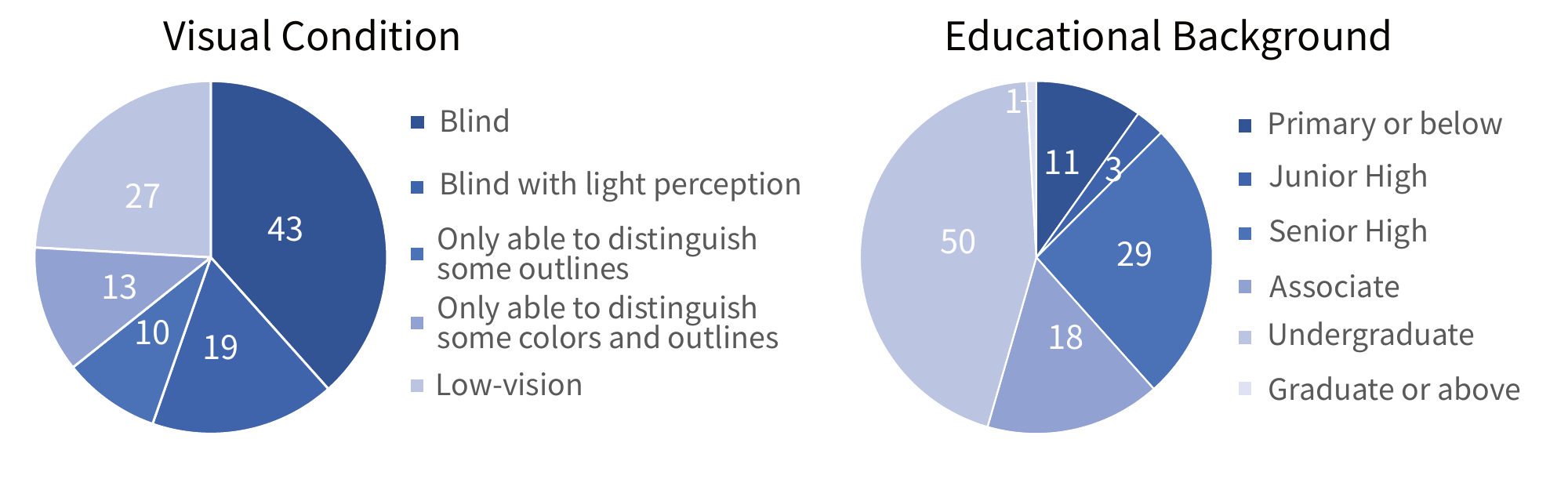}
  \caption{Visual condition and educational background of the questionnaire recipients.} \label{Fig: Question Result}
  \Description{Visual condition and educational background of the questionnaire recipients.}
\end{figure}

We distributed the online questionnaire through the \textit{WJX} platform\footnote{\textit{WJX} platform: \url{https://www.wjx.cn/}} using a \textit{Snowball Sampling Method} within BVI communities, aimed to gather background information and initial insights into the spatial perception challenges faced by BVI individuals. \chadded{The BVI used the screen reader} \footnote{\chadded{For example, Android offers the official talkback function: \url{https://support.google.com/accessibility/android/topic/3529932}}} \chadded{to obtain text information and fill out the questionnaire independently.} Overall, we have received 112 valid responses, including 44 from individuals who are completely blind, 19 from those with light sensitivity, and 23 from participants with partial sight,  as shown in Figure \ref{Fig: Question Result}.



\subsubsection{\textbf{Semi-structured Interviews.}}

To gain a deeper understanding of the mental state, needs, and challenges faced by BVI individuals in relation to scale perception, we conducted a series of semi-structured interviews. These interviews complemented the open-ended questions in our questionnaires, providing BVI participants with opportunities to freely share their personal experiences and perspectives on scale perception.

We selected 12 participants from our questionnaire respondents, ensuring a diverse mix in terms of living areas, genders, ages, and visual conditions (6 males and 6 females, ages ranging from 22 to 43, Mean = 30.08, SD = 7.09), as shown in Table \ref{tab:1 Demog}. To minimize potential biases, half of the interviewees self-identified as having strong scale perception abilities, while the other half perceived their abilities as weak. The interviews, which lasted between 50 to 90 minutes, were conducted either one-on-one or in small groups of 3 to 4, using a mix of online and in-person methods for accessibility.

\begin{table*}   
\footnotesize
\setlength{\tabcolsep}{3.5pt}
   \renewcommand\arraystretch{1.4}
  \caption{Self-reported demographic information of the selected 12 interviewees}      \label{tab:1 Demog}
\centering
\begin{tabularx}{\textwidth}{ccccp{25mm}ccccc}
    \toprule
    \textbf{No.} & \textbf{Age} & \textbf{Sex} & \textbf{Area} & \textbf{\makecell{Visual\\Condition}} & \textbf{Job} & \textbf{Hobbies} & \textbf{\makecell{Research\\Methods}} &
    \textbf{\makecell{Education}} & \textbf{\makecell{Residential\\Status}} \\    \midrule
    P1 & 28 & M & urban & congenital low vision & pianist & \raggedright handcraft, reading & Q, I, D & bachelor's & with colleagues \\
    P2 & 28 & M & urban & congenital blindness & doctor & \raggedright hand knitting & Q, I, D & bachelor's & with wife\\
    P3 & 43 & F & urban & acquired blindness & braille editor & \raggedright reading, dancing & Q, I, D & bachelor's & alone \\
    P4 & 22 & F & rural & congenital blindness with light perception & unoccupied & \raggedright origami, reading & Q, I, D & bachelor's & alone\\
    P5 & 28 & M & urban & congenital blindness with light perception & piano tuner & \raggedright playing games & Q, I, D & bachelor's & with roommates \\
    P6 & 27 & F & urban & congenital blindness & braille editor & \raggedright voice book listening & Q, I, D & bachelor's & with family \\
    P7 & 40 & M & rural & congenital blindness with light perception & massagist & \raggedright music listening & Q, I & bachelor's & with colleagues \\
    P8 & 35 & F & rural & acquired blindness & unoccupied & \raggedright music listening, reading & Q, I & junior high's & with family\\
    P9 & 24 & M & urban & congenital blindness & unoccupied & \raggedright reading, playing games & Q, I & bachelor's & with roommates  \\
    P10 & 38 & F & rural & congenital blindness & massagist & \raggedright origami, reading & Q, I & senior high's & with colleagues \\
    P11 & 25 & F & urban & congenital blindness with light perception & massagist & \raggedright origami, reading & Q, I & bachelor's & with family \\
    P12 & 23 & M & urban & acquired blindness & unoccupied & \raggedright playing chess, reading & Q, I & bachelor's & with roommates \\
  \bottomrule
  \multicolumn{10}{l}{For research methods, Q indicates \textit{Questionnaire}, I indicates  \textit{Interview}, and D indicates  \textit{Door-to-door Study}.}
\end{tabularx} 
\end{table*}


The interviews were conducted within two weeks. Initially, participants were asked to provide a brief recap of their basic information (5 minutes). They then described the sizes and angles of everyday objects, allowing us to assess their actual scale perception abilities (25 minutes). Participants also discussed scenarios where they encountered challenges with scale information, sharing any strategies they used to overcome these challenges (40 minutes). Finally, they expressed their expectations and preferences regarding methods to train and improve their scale perception skills (20 minutes). All interviews were recorded with the participants' consent.

\subsubsection{{\textbf{Door-to-door Study}}}

Though we interviewed BVI individuals about how they perceive object scales, it is challenging to verify the truthfulness of their subjective descriptions, especially with personal items. To address this, we conducted a \textit{Door-to-door} (DTD) study with six local BVI participants from our interview group (see Table \ref{tab:1 Demog}), visiting them one by one in their homes or offices. This approach aimed to provide a more accurate and personal understanding of how they perceive object sizes in their daily surroundings.



\begin{figure}[h] 
  \centering
  \includegraphics[width=0.8\linewidth]{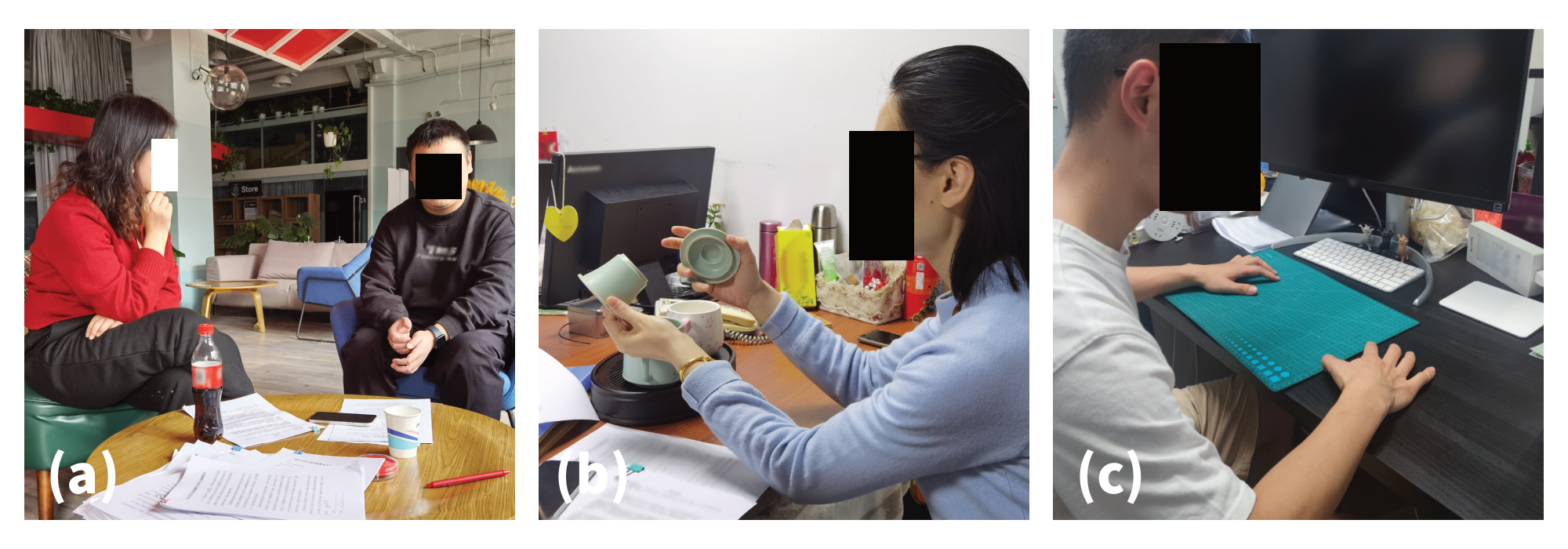}
  \caption{We are doing the door-to-door research. Those with black rectangles covering their faces are the blind or low-vision. (a) Talking about his own situation. (b) Describing the sizes of objects around her. (c) Trying to illustrate a size on a mat with scales.} \label{Fig: Pre research}
  \Description{We are doing the door-to-door research. Those with black rectangles covering their faces are the blind or low-vision. (a) Talking about his own situation. (b) Describing the sizes of objects around her. (c) Trying to illustrate a size on a mat with scales.}
\end{figure}

The process, as described in Figure \ref{Fig: Pre research}, began with a review of information from earlier interviews, supplemented with any new insights. Next, participants were guided through both descriptive and practical tasks. They were asked to describe objects, like their favourite cups, with as much accuracy as possible and to use natural gestures to convey the objects' dimensions. During these activities, we carefully observed and noted their speech patterns and physical actions, providing assistance whenever challenges arose. This approach also allowed us to engage in discussions about any particularly interesting observations made in real time. Explicit consent was obtained from all participants, including taking observation notes, photographing the environments and objects, and recording key conversations.


\subsection{Findings} \label{Cap: Findings and Discussions}


During the data analysis, initial statistical reports of online questionnaires were automatically generated by \textit{WJX} platform. All interview sessions were audio-recorded and subsequently transcribed for detailed examination. Complementing these methods, observational notes were taken to capture the perspectives of BVI participants during the DTD phase. Employing \textit{Thematic Analysis} \cite{Braun2006Using}, we open-coded both interview transcripts and notes, iterating collaboratively with the full research team to refine and synthesize insights. This integrated approach, merging quantitative questionnaire data with qualitative interviews and observational findings, provided a comprehensive understanding of the BVI community's daily challenges and strategies in relation to small and medium scales. Main findings are summarised below.


\subsubsection{Main barriers.} \label{Cap: Main barriers} Here, we identify four main barriers limiting the cultivation of scale perceptions among BVI individuals:

\textbf{(1) Social stress and isolation.} Precise training in measurement accuracy is often confined to specialized schools, making it inaccessible to the broader BVI population. Our questionnaires indicate that 64\% (72 out of 112) of BVI individuals seek help from family or friends, but support is inconsistent, sometimes delayed for hours or days, or even unresponsive. This reliance can create social pressure, with about 29\% (33 out of 112) of participants considering abandoning their efforts due to such pressures. Additionally, introversion in some BVI individuals further limits their access to information about scales and assistance (P4, P8, P10, P25).

\textbf{(2) Inefficient cumulation of scale perception knowledge.} Efficient information input channels play a crucial role in fostering spatial perception among BVI individuals. Through the interview, we noticed that BVI individuals construct {\itshape``knowledge repositories''} through memorization of object sizes, enabling them to comprehend their environment. However, challenges arise in accessing information, and disparities in perception compared to sighted individuals hinder the efficient accumulation of these repositories. In order to meet the needs of dimensional information acquisition, 60 out of 112 individuals reported having a braille ruler at home for measurements in the questionnaire, yet its bulkiness poses challenges for portability. Moreover, interviews revealed that communication gaps between BVI and sighted individuals, who often rely on vague verbal descriptions, contribute to inefficient information input (P6). Sighted individuals may use imprecise terms, hindering BVI understanding. For example, participant P3, who works at the Museum for the Blind, mentioned a discussion about the design of the hall layout. When a sighted coworker suggests that {\itshape``This (ad page) is too wide for the lobby, let's opt for a smaller one''}, it can pose difficulties for BVI individuals, especially those who are totally blind, as they may struggle to grasp the meaning of {\itshape``this''} and imagine the size of the {\itshape``lobby''}, which do not contribute to the construction of their ``knowledge repository''. Clear and specific descriptions, include specific names, dimensions, and analogies such as {\itshape ``This mug is 15 centimeters high''} or{ \itshape ``The lid is as wide as the palm of your hand''} enhance information accessibility.


\textbf{(3) Gap between \itshape"knowing" and \itshape"doing".} Despite adding specific sizes and names to conversations, individuals with BVI may not fully grasp the meaning of sizes. From interviews and DTD observations, we noticed a gap between ``knowing'' and ``doing'' in the BVI community, particularly evident in activities like Tai Chi, dance, or yoga using angle descriptions. The questionnaire indicates 65.49\% (74/112) of the BVI expressing uncertainty about the accuracy of their spatial scale understanding, with an additional 10.71\% (12/112) noting for their understanding of spatial scale. Another 12 chose 'fuzzy concepts and frequent mistakes' or 'not received specific training'. In physical demonstrations during DTD research, even if BVI understood concepts like {\itshape``45 degrees is half of a right angle''}, imprecise movements were common. It is also difficult to describe the size of an object based on independent exploration. Notably, some showed exceptionally accurate movements, like P2 demostrating 15 cm, who had meticulously measured the length of his thumb and pinkie when they were tight and had memorized the associated muscular sensation. Even so, he did not know of other dimensions that he had not actively practiced In the correction process, we found that just informing a BVI person of accuracy wasn't enough. We needed to  guide them to experience the sensation of the target size with detailed instructions. However, even if participants accurately achieved the target scale with our guidance, reproducing it consistently could lead to challenges were challenging. Achieving a strong linkage between scale information and body state requires a gradual and intensive training process, not achievable overnight. 

\textbf{(4) Limited exploration opportunities.}
Blind individuals often face limited opportunities to explore the environment around them, particularly if they lack adequate support. When coupled with social pressures, self-doubt, introversion, and ineffective information feedback, as mentioned earlier, these factors collectively constrain the BVI from fully exploring their surroundings. Consequently, their opportunities to comprehend spatial relationships across diverse contexts are restricted, impeding the accumulation of life experience related to scale perception{, as revealed in interviews and DTD research (P1-P6)}.


\subsubsection{Current tactics} \label{Cap: Natural Methods of BVI People}

\begin{figure}
  \centering
  \includegraphics[width=0.8\textwidth]{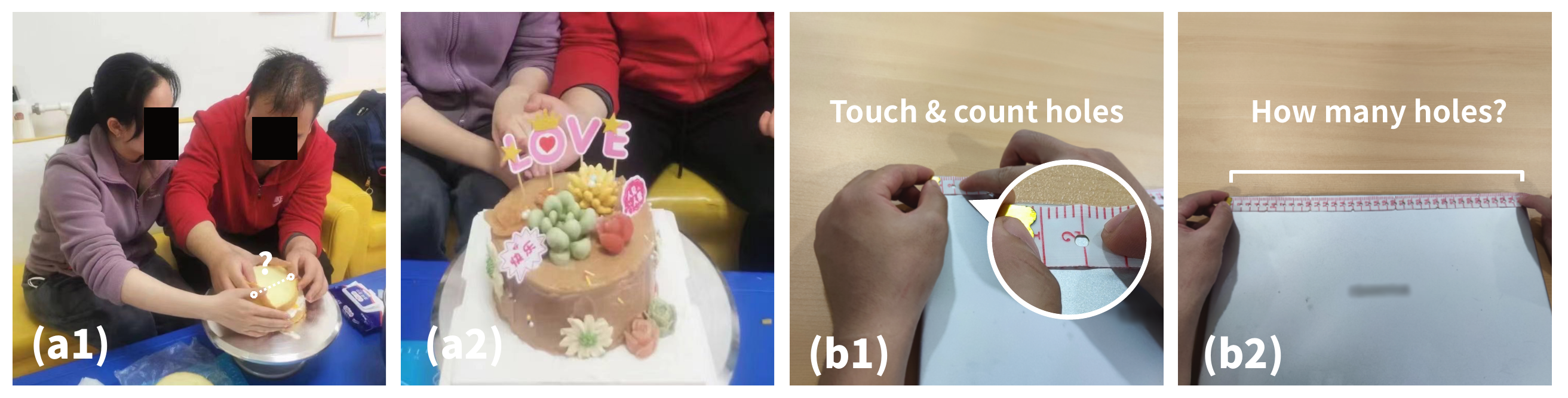}
  \caption{The BVI face many realistic problems related to scales in their daily life and try to solve them although it is time-consuming and laborious. 
 (a1)and (a2) show that a blind couple uses their hands to feel the size of the cake. (b1) and (b2) show that a blind person uses a braille ruler to measure the laptop by touching and counting holes.} \label{Fig: Realistic Problems}
  \Description{The blind and low-vision face many realistic problems related to scales in their daily life.}
\end{figure}

While visual disabilities pose challenges (Figure \ref{Fig: Realistic Problems}) for the BVI  in handling small and medium scale issues \cite{Children1, Children3,revesz1950psychology}, they have pursued various tactics. Here, we summarize their experiences and analyze their limitations.

\textbf{{(1) }Find an intermediary.}
When describing scale information, the majority (85/112) of BVI individuals indicated that they prefer to use familiar objects as {\itshape ``intermediaries''} for communication. In interviews, P2 said, {\itshape{\itshape``Remember the length of the phone to measure the width of the table''}}, {and} P3 said, {\itshape{\itshape``This cup is a little bigger than the one we used last time''}}. This method is suitable for interlocutors with similar life circumstances and intersections, but if the two sides of the dialog cannot find a suitable intermediary medium, the communication is prone to misunderstanding (P6).

\textbf{{(2) }Use {the} body as a ruler.}
Our comprehensive research revealed that BVI individuals with robust spatial perception often employ their limbs as ``measuring tools''. They memorize the mapping relationship between scales and specific body parts' sizes, such as palm size, height, forearm length, and even fingernail width. In the questionnaire, 42 respondents mentioned this method, and nearly all interviewees endorsed it, as P4 said, {\itshape{\itshape`` The base of my nose to my chin is 5cm, which I had specially measured when I studied at college. I use it to estimate the size of other items when necessary.''}} Also, they said hands are important rulers to explore, which was consistent with \cite{dim2014designing}. {Besides, different gestures involving muscular memories of fingers, hands, limbs, and torso, are also recognized as important tools when exploring. Some typical gestures of them are observed during the DTD research, and summarized in the left top ``Natural Gestures'' part of Figure \ref{Fig: System Design}.} Simultaneously, they acknowledged that their accumulation in this regard is often quite limited, typically consisting of only vague impressions. Additionally, the little attention to muscular memory hampers understanding scale information (Almost all). For example, the ones with poor performance tend to describe all sizes around 8cm which is the distance between their thumb and forefinger when their hand is in a relaxed state, neglecting the {tiny} sensory differences between muscles, thus making themselves always unable to clearly distinguish and understand these sizes.



\textbf{{(3) }Memorize the mapping of values and objects.}
Focusing on accumulating life experiences stands as another vital approach to shaping spatial perception. Despite the challenge of translating abstract values into physical space, many individuals (almost all) make a deliberate effort to memorize specific measurements of commonly used items, such as the pot lid at home is 36 cm, the table is 180 cm wide, and the distance from the bed to the wall is 140 cm etc., to facilitate daily communication requirements. This method is often combined with the method of using the body as a ruler, and {so} the accumulation of this knowledge base is limited by the {same} limitations mentioned above.




\subsubsection{Classification of Small and Medium Scale Demands} \label{Cap: Classification of Demands}

{We summarized the scale range and precision requirements required in the lives of BVI. This paragraph is based on DTD research and interviews.}

\textbf{(1) The range of scales.}
We find that the small and medium-size demands of BVI individuals primarily revolve around item size assessment and positional relationships. When it comes to angles, their focus is primarily on orientation judgments based on self-reference. We summarized the top 5 demands scenarios with typical examples exhibited in Table \ref{tab2:Top 5 scenarios}. {For the five typical natural gestures we summarized, their scale ranges are shown in Table \ref{tab3:Gesture model}.}

\begin{table*}   
\footnotesize
\centering
   \renewcommand\arraystretch{1.5}
  \caption{Top 5 scenarios that require spatial scale found in pre-research} 
  \label{tab2:Top 5 scenarios}
  \begin{tabularx}{\textwidth}{p{2.5cm}clX}
    \toprule
    \textbf{Scenarios} & \textbf{\makecell{Mentions in\\ Questionnaire Results}} & \textbf{\makecell{Interviewees\\Mentioning}} & \textbf{Typical Examples} \\
    \midrule
    \raggedright Communicate with others, especially about topics of scales & 52 & Almost all & {\itshape{``When talking with my blind friends, they may describe a size according to their own item. But I don't know that item so it is difficult for me to understand.''}} \\
    \hdashline
    \raggedright Select and purchase goods & 40 & Almost all & {\itshape{``I have a hard time understanding items' sizes according to the product description in online shopping. This results in very limited items that I can purchase on my own.''}}\\
    \hdashline
    \raggedright Arrange the surroundings & 30 & P8, P10 & {\itshape{``I hope that I can arrange the items in my room systematically based on my understanding of the size of different items.''}} \\
    \hdashline
    \raggedright Do daily life tasks or hobbies & 28 & Almost all & {\itshape{ ``I usually make handicrafts in my spare time. It was really difficult for me at first for my few perceptions about scales.''}}\\
    \hdashline
     Position self in the space & 21 & P1, P10 & {\itshape{``When visiting large shopping malls, I hope to use my own position as a reference frame and construct walking paths by asking passers-by, so as to find the commodity area I want to go to.''}}\\

  \bottomrule
\end{tabularx}
\end{table*}

\textbf{(2) Precision needed in different scenarios:}
The BVI's needs for precision vary from scenario. Sometimes they would like to know exact values of sizes or angles, such as when buying a table and wanting to know its size. And the more accurate value they get, the better they will handle the shopping problem and even learn about the related sizes. These scenarios are usually related to measurements or mapping from objects to scales. However, exact precision may not be necessary for other scenarios, especially when it is from scales to objects or movements. For example, when they get the guide that they need to move forward one meter to reach a trash can, they do not need to walk exactly one meter without any difference (and it is almost impossible), where tolerances should exist.

\subsubsection{Feedback for Dual-channel based Instructions}\label{Cap: Needs for Dual-channel based Instructions} 

{Assisting individuals with BVI during DTD research highlighted the importance of providing feedback through multiple channels. Effective feedback integrates various sensory cues and technical guidance.} Auditory signals, conveying text and sound cues, offer valuable guidance. But hearing and understanding speeches often cost some time, and so decrease the effects of real-time instructions. Instead, if we ``communicate'' with the BVI through a sense of touch, like holding his or her hand to avoid moving farther or tapping his or her left shoulder, he or she can usually respond quickly, stopping right there to feel the size or angle. Thus, instructions based on dual-channel and combining touch and sense are important for good teaching effects.

{We also find that instruction texts need to strike a balance between specific and unspecific words, considering the BVI's challenge with precise values. While unspecific instructions like ``a little bit'' are easier to understand and execute, precise values (e.g., ``30 cm'') strengthen the connection between body movements and scale concepts. For instance, P2 suggested in the door-to-door task that specific scale reminders aid in consolidating mastered scales, stating {\itshape ``Reminding me of the specific scale and the degree of adjustment can help me consolidate the scale I have mastered.''}}

 \begin{table}  
\footnotesize
\renewcommand\arraystretch{1}
\caption{Gesture model of scale perception in blind people} \label{tab3:Gesture model}
   \setlength{\tabcolsep}{5.3mm}{
  \begin{tabular}{lll}
    \toprule
    \  \textbf{Body parts}  & \textbf{Gestures} & \textbf{Range} \\
    \midrule
    \textbf{One finger} & Move a distance finely & 0 \textasciitilde 12cm  \\
    \textbf{Two fingers} & Open and close finely to adjust the distance between fingers & 0 \textasciitilde 12cm \\
    \textbf{One hand} & Move a distance & 0 \textasciitilde 120cm  \\
    \textbf{Two hands} & Open and close to adjust the distance between hands & 0 \textasciitilde 120cm  \\
    \textbf{Body} & Rotate to change orientation & 0 \textasciitilde 360deg \\
  \bottomrule
\end{tabular}}
\end{table}

 \begin{figure*}
     \centering
     \includegraphics[width=\textwidth]{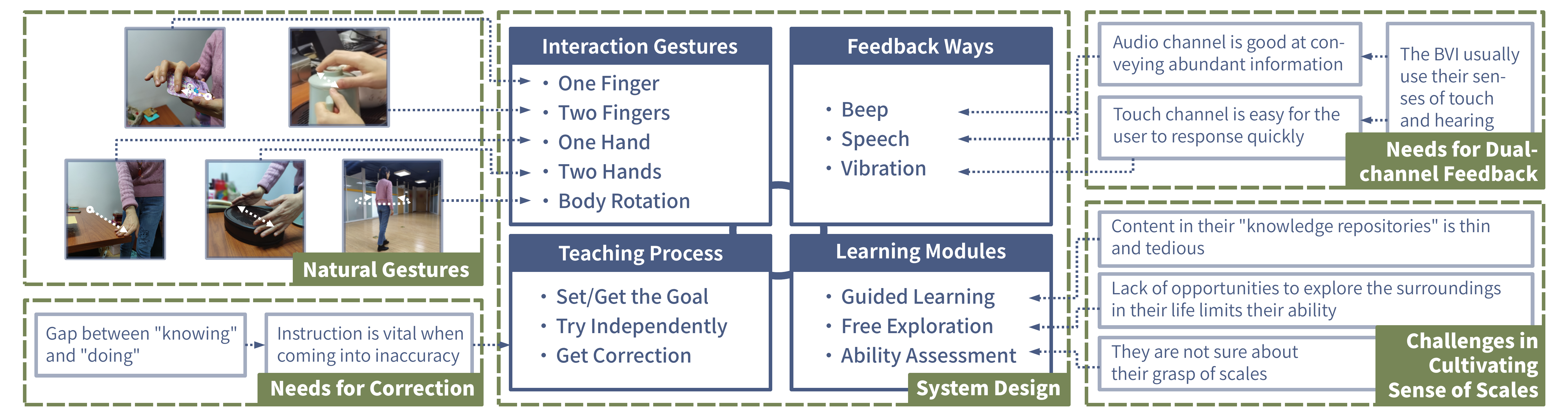}
     \caption{The system design of AngleSizer inspired by preliminary research, including the interaction gestures, feedback mechanisms, teaching processes, and learning modules.} \label{Fig: System Design}
 \end{figure*}

\section{SYSTEM DESIGN} \label{Cap:System Design}

\begin{figure*}[h] 
  \centering
  \includegraphics[width=\textwidth]{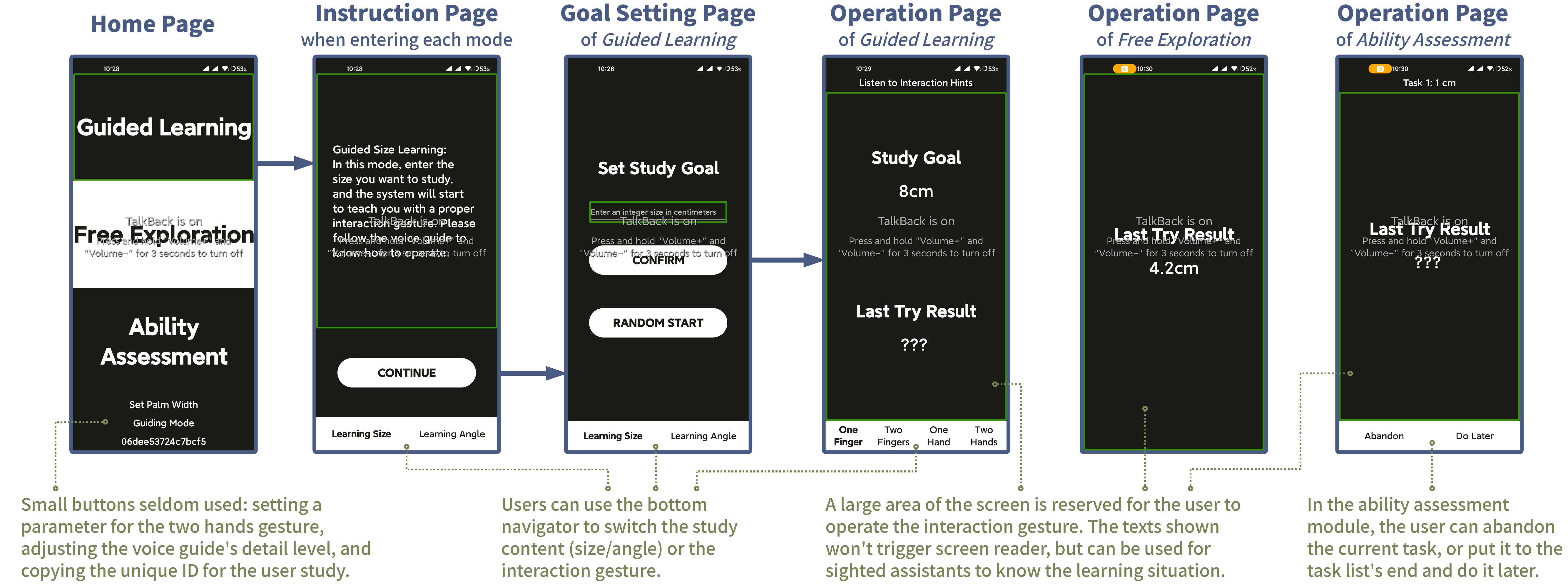}
  \caption{\chadded{Screenshots of AngleSizer's. The accessibility service is on, and the green-boarded box on each page is the active UI element of the screen reader.}} \label{Fig: UI Interface}
  \Description{Screenshots of AngleSizer's some pages. The accessibility service is on, and the green-boarded box on each page is the active UI element of the screen reader.}
\end{figure*}

Inspired by the formative study in Section \ref{Cap:Preliminary research}, Figure \ref{Fig: System Design} illustrates the design of AngleSizer, a smartphone-based teaching assistant. It features natural gesture-based teaching, comprehensive feedback, meticulous teaching process management, and flexible learning modules. \chadded{Additionaly, as shown in Figure~\ref{Fig: UI Interface}, the user interface is simple and accessible to BVI users, facilitating easy navigation with the phone's native screen reader.}

\subsection{Interaction Gestures Design} \label{Cap:4.1 Interaction Gestures}
\begin{figure*}
  \centering
  \includegraphics[width=.99\textwidth]{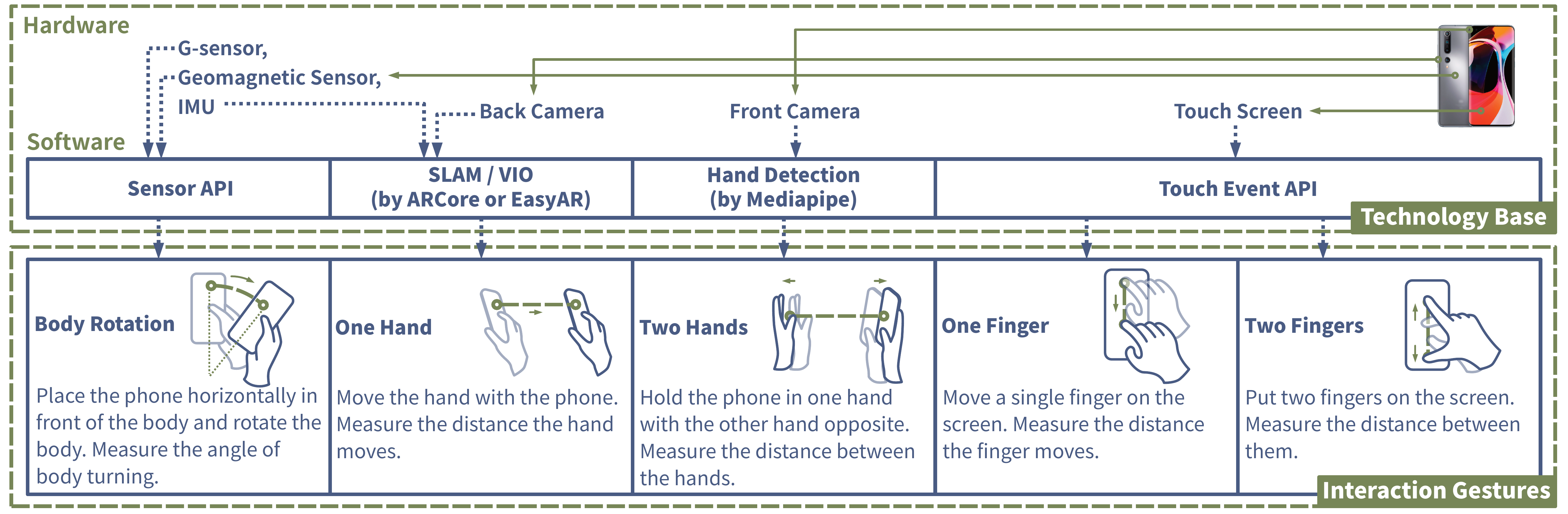}
  \caption{Utilizing the hardware and software features of smartphones, AngleSizer supports five interaction gestures to show sizes and angles.} \label{Fig: Technology and gesture}
  \Description{Utilizing the hardware and software features of smartphones, AngleSizer supports five interaction gestures to show sizes and angles.}
\end{figure*}

Based on common measurement gestures used by the BVI (Table \ref{tab3:Gesture model}), {and their habitual uses of hand gestures with mobile devices \cite{dim2014designing},} we designed five interaction gestures: {\itshape One Finger}, {\itshape Two Fingers}, {\itshape One Hand}, {\itshape Two Hands} and {\itshape Body Rotation}. The bottom part of Figure \ref{Fig: Technology and gesture} demonstrates the use of these gestures. For finger gestures, users simply place and move their fingers on the screen. For other gestures, they press the screen to initiate interaction. All gestures are designed with a sufficiently large operational area on the screen to ensure ease of use.

While the system can calculate distances or angles with high precision, our findings in Section \ref{Cap: Classification of Demands} suggest that BVI users often require varying levels of accuracy depending on the context. For tasks requiring high precision, the system provides measurements with millimeter accuracy for finger gestures, centimeter accuracy for hand gestures, and angle measurements accurate to 1 degree. 
For less precise requirements, the system introduces a tolerance based on research into size and angle perception \cite{Worchel1951SpacePA, VisualHapticAnglePerception, IndenpendenceSizeDistance, cutting1995perceiving, Withagen2004, Jones2012Accuracy}, accommodating the practical needs of BVI users.



\subsection{Feedback Design} \label{Cap: 4.2 Feedback Design}


In accordance with the needs for dual-channel based instructions outlined in Section \ref{Cap: Needs for Dual-channel based Instructions}, the system employs a combination of audio and tactile feedback. Audio feedback includes beeps and speech, while tactile feedback is provided through vibrations. These methods have been proven effective in enhancing user interaction as highlighted in studies such as Tactile Compass \cite{TactileCompass}. Specifically:

\textbf{Beep feedback.} The system utilizes various distinct beeps to guide user interactions. A crisp beep indicates a correct gesture, while a warning beep signals an error. For instance, the two-hand gesture requires one palm to remain within the front camera's view. The system emits a correct beep when the camera detects the palm and an error beep if the palm exits the view, aiding users in adjusting their hand positions. \chadded{Similarly, beeps help users maintain the phone as parallel to the ground as possible, enhancing the accuracy of rotation angle measurements.}

\textbf{Speech feedback.} Speech feedback provides detailed instructions and results. It offers step-by-step guidance during gesture performance and reports the outcome and its precision upon completion. The speech content is crafted to utilize both specific and general expressions, adhering to the principles discussed in Section \ref{Cap: Needs for Dual-channel based Instructions}.

\textbf{Vibration feedback.} Vibration feedback complements speech by enriching the user experience. It not only confirms the initiation of a measurement with a spoken "the measurement started" but also through a tactile vibration to ensure the user is aware. Moreover, vibrations serve as a continuous guidance tool, especially useful in the correction phase of the teaching process, detailed in Section \ref{Cap: 4.3 Teaching Process Design}.

\subsection{Teaching Process Design} \label{Cap: 4.3 Teaching Process Design}


As highlighted in Section \ref{Cap: Main barriers}, the gap between "knowing" and "doing" necessitates a teaching process that extends beyond simple correctness feedback to provide more hands-on instruction. Our system is structured around three phases: \textit{setting or identifying a goal}, \textit{independent practice}, and \textit{correction}. 

Initially, the user sets a target value for sizes or angles. The system then allows the user to attempt to reach this target independently and subsequently checks if the outcome is within the predefined tolerance (as detailed in Section \ref{Cap:4.1 Interaction Gestures}). If the attempt is not successful, the system enters the correction phase, using vibrations to guide the user toward the correct positioning. This tactile feedback helps users recognize the correct scale or alignment by feeling, thereby memorizing the correct muscular movements and values. After this correction, the user is prompted to try again independently to reinforce learning.

\subsection{Learning Modules Design} \label{Cap:4.4 Learning Modules Design}

The system includes various learning modules designed to address specific learning challenges identified for BVI users (see Figure \ref{Fig: Teaching process}). Specifically: 

\begin{figure*} 
    \centering
    \includegraphics[width=\linewidth]{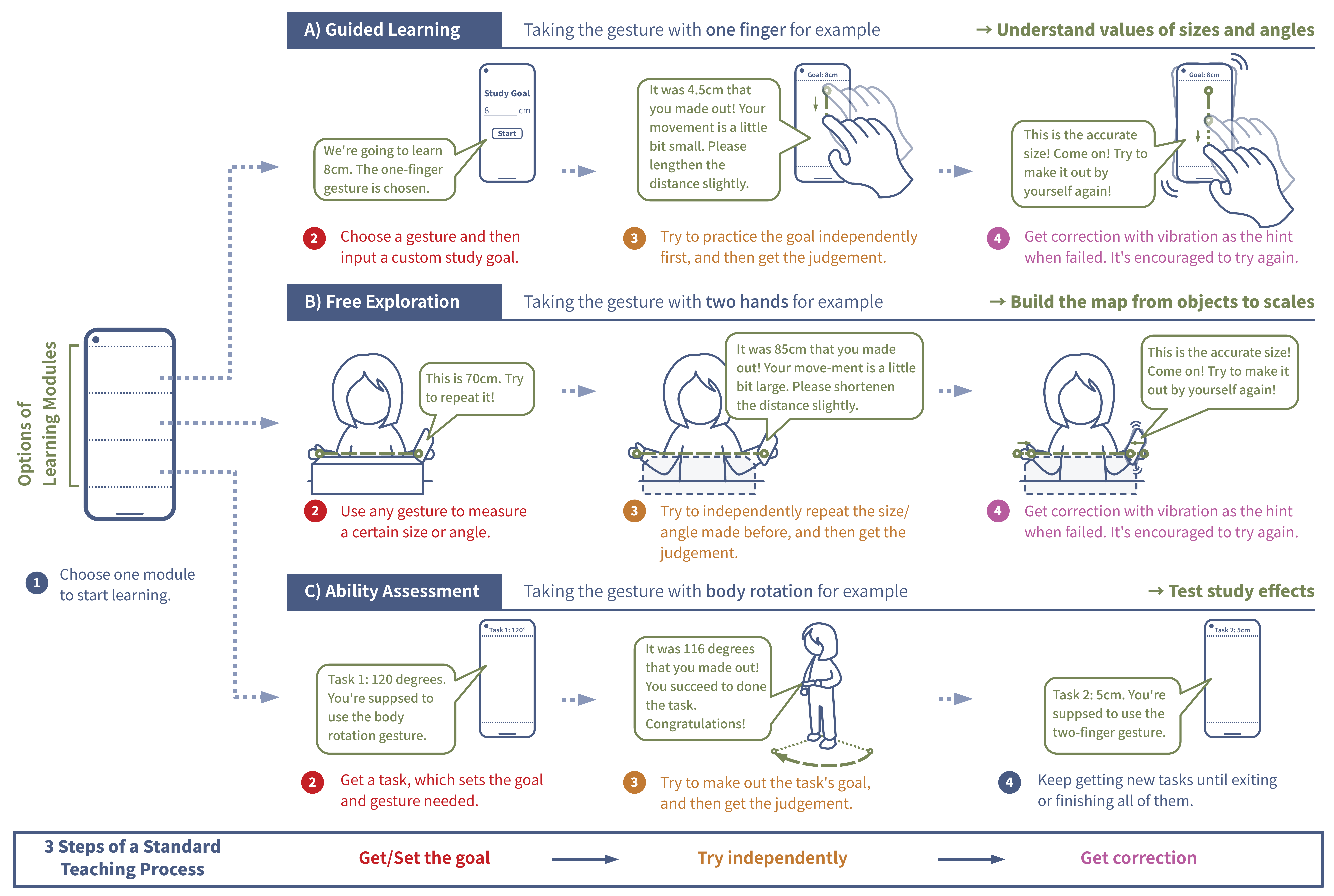}
    \caption{AngleSizer offers three learning modules modifying the standard teaching process, to solve different learning problems respectively.} \label{Fig: Teaching process}
    \Description{AngleSizer offers three learning modules modifying the standard teaching process, to solve different learning problems respectively.}
\end{figure*}

\textbf{Guided Learning.}
In this module, users actively engage by setting a specific size or angle they wish to learn and selecting an appropriate gesture. The system provides step-by-step instructions for executing the gesture, such as \textit{"Hold on for 3 seconds, and move your finger when the phone vibrates"}, ensuring users learn the correct execution method right from the start.

\textbf{Free exploration.} 
This module allows users to learn through unstructured exploration, integrating learning with daily activities and overcoming the challenge of limited exploratory opportunities. Users can perform any gesture to interact with their environment; the system identifies these gestures and provides immediate feedback on the values achieved, turning each action into a learning opportunity. Then, the value is considered as the study goal. The user is encouraged to repeat it as the independent try phase {and get correction later when necessary. This method helps users develop a mental map linking physical objects with their dimensions, reinforcing knowledge through practical experience.

\textbf{Ability Assessment.}
In this module, users are tested on their ability to accurately produce a size or angle using a specific gesture. This module combines tasks from the previous modules and introduces common measurements relevant to daily life. The system evaluates the accuracy of each task and confirms whether the results fall within an acceptable tolerance range. Unlike other modules, there is no correction phase here, emphasizing the assessment of the user's independent capability.

\subsection{Technical Description} \label{Cap: Technology Base}

\textbf{Technology base of gestures.}}
Our system prototype is implemented on Android phones and leverages various hardware and software capabilities specific to each type of interaction gesture, as depicted in Figure \ref{Fig: Technology and gesture}. For gestures involving fingers, we utilize the \textit{TouchEvent} API to capture the coordinates of the fingers on the screen in pixels and calculate the actual distance using the screen's DPI (dots per inch) property. The one-hand gesture employs the phone's back camera and IMU, integrating AR frameworks' motion tracking capabilities to determine the translation of the camera's coordinate system relative to the world's coordinate system, thereby calculating the distance between different phone positions. The two-hand gesture uses \textit{MediaPipe} to recognize palms within the front camera's view and estimates the distance relative to the phone based on a pre-entered palm width and its displayed size. For measuring angles through body rotation, we use the \textit{Sensor} API, which utilizes data from the G-sensor and geomagnetic sensor to calculate the \textit{Euler} angles of the phone, accumulating these over time to determine the total rotation angle.

To address potential noises and errors, and to refine our calculations, we have implemented several optimization strategies:

\chadded{(1) \textit{Coordinate Stability}: To counteract sudden leaps in coordinates that typically occur during SLAM's coordinate system reconstruction, we've set a parametric threshold. If the length of the change vector between coordinates in consecutive frames exceeds this threshold, the vector is resized accordingly.}

\chadded{(2) \textit{Jitter Reduction}: Acknowledging the jitter that can occur when users try to hold their position at the end of an action, we calculate the final reported position as the average over the last few moments to ensure stability and accuracy.}

(3) \textit{Palm Parallelism Monitoring}: Distance measurement using the Two Hands gesture depends on precise palm width measurement, where significant errors arise when palms are not parallel to the smartphone screen (especially when the palm rotates along the longitudinal arch). To minimize these errors, we track the depth coordinates relative to the wrist node (Z values) of multiple points along the distal transverse arch to detect significant palm rotations. Upon detecting a substantial rotation, a warning beep reminds users to adjust their palms to a position parallel to the smartphone screen.

\subsubsection{\textbf{Workflow of the system.}} \label{Cap: Work Flow}
\begin{figure*}[h] 
  \centering
  \includegraphics[width=\textwidth]{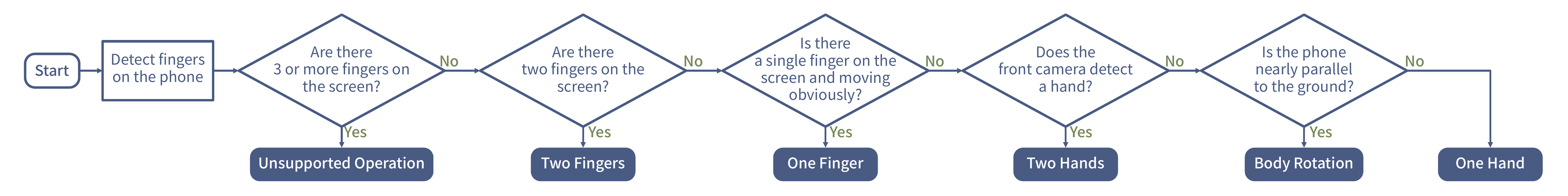}
  \caption{{The logic of how the system determines the activated gesture in the \textit{Free Exploration} module.}} \label{Fig: Gesture logic}
  \Description{The logic of how the system determines the activated gesture in the free exploration module.}
\end{figure*}

When the user enters a learning module, the system will wait for him or her to start a gesture. {In the \textit{Guided Learning} and \textit{Ability Assessment} modules, the system only checks when the user's action fits the requirements of the chosen gesture (described in Section \ref{Cap:4.1 Interaction Gestures}); while in \textit{Free Exploration}, it will determine the activated gesture according to the logic shown in Figure \ref{Fig: Gesture logic}.}

\begin{figure*}[h] 
  \centering
  \includegraphics[width=\textwidth]{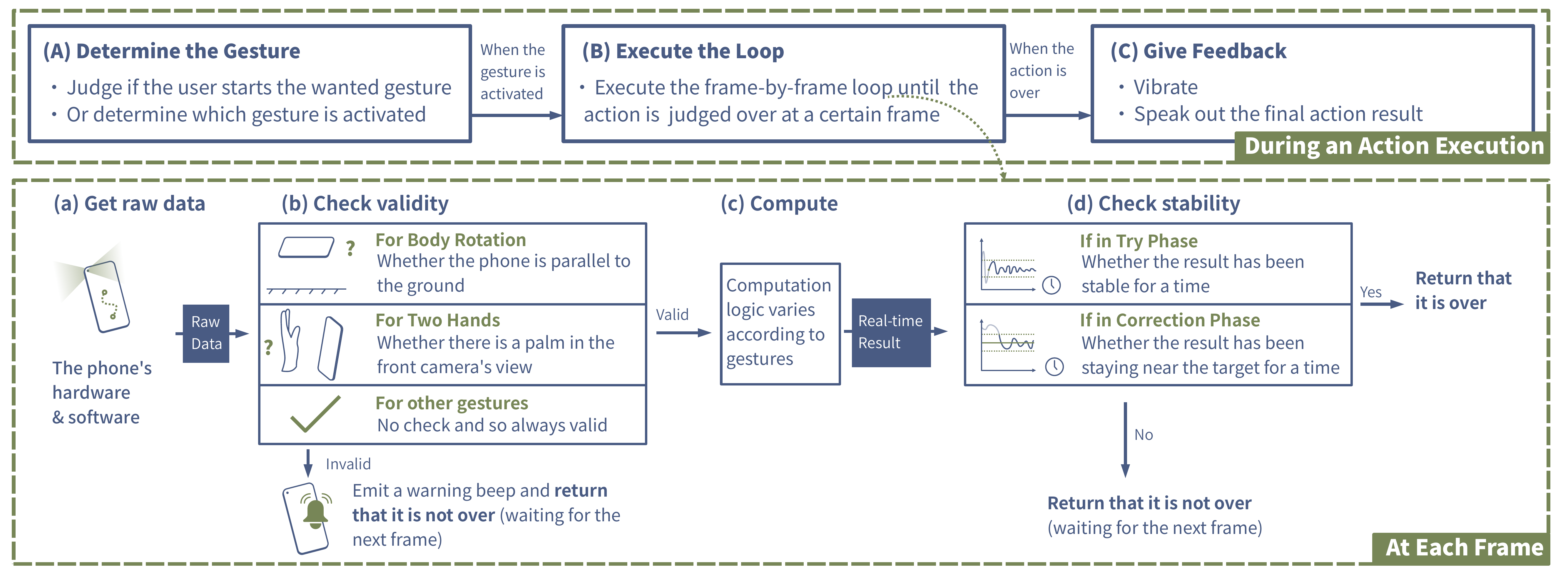}
  \caption{When AngleSizer is running, it will first judge the activation of the interaction gesture. Once a gesture is activated, the system will continue getting raw data, checking validity, computing, and checking stability works at each frame. When a certain frame judges that the action is over during checking stability, the system will break the loop and give feedback with vibration and speech.} \label{Fig: Frame process}
  \Description{When AngleSizer is running, it will first judge the activation of the interaction gesture. Once a gesture is activated, the system will continue getting raw data, checking validity, computing, and checking stability works at each frame. When a certain frame judges that the action is over during checking stability, the system will break the loop and give feedback with vibration and speech.}
\end{figure*}

For any gesture, once it is activated, the system will record the initial position or orientation, and execute a frame-by-frame loop, doing check and calculation works in real-time as Figure \ref{Fig: Frame process} shows. 
When the stability check at some frame is passed, the system will break the loop and give feedback, speaking out the measurement result (in the independent try phase), or encouraging the user to remember the feeling and try again (in the correction phase). All the above is the flow of a single teaching process, and the system will wait for another teaching process, such as the correction after the independent try.

\subsection{\chadded{System's Accuracy}\label{Cap:System's Accuracy}}

\chadded{Morar et al. evaluated ARCore and obtained an average position error of 0.16m \cite{Morar2020Evaluation}. Kuhlmann et al. found that the mean deviations for static pitch and roll values a smartphone calculated are 0.05° and 0.20° respectively, and that the sensor accuracy varies from device to device \cite{Kuhlmann2021}; however, they did not test the yaw values nor the angle change during a dynamic rotation process, which are mainly used in our system's angle measurement.}


\chadded{Considering the technical constraints mentioned above and to ensure that AngleSizer users can learn the correct scale, we evaluated the accuracy of the system with two measurements: the system's \textbf{\textit{inherent error}} and its \textbf{\textit{operational error}}.
The system's inherent error refers to its precision relative to the objectively true scale values (measured using a ruler or protractor). The system's operational error refers to the measurement error in actual usage scenarios, considering both the system's inherent errors and those caused by improper operation (such as hand jitter) by BVI users.}

\chadded{To measure the \textit{inherent error}, two experimenters with normal vision (2 males, aged 22 and 23 respectively) replicated target scales using AngleSizer with the assistance of precise measuring tools (rulers and protractor). Each experimenter replicated each target value five times per gesture. During the measurement process of \textit{One Finger} and \textit{Two Fingers}, the experimenters placed a ruler with precise markings on the phone screen and slid their fingers to measure with visual and tactile assistance. Similarly, for \textit{One Hand} and \textit{Two Hands} measurements, a tape measure was used as a reference, and the phone or palm was moved strictly along the ruler's markings to measure distance. \textit{Angle} measurements were conducted by rotating the phone along the printed protractor's markings. Table~\ref{Sighted Measure Accuracy} illustrates the inherent errors of AngleSizer, with an average relative error of 1.22\% for on-screen gestures (\textit{One Finger}, \textit{Two Fingers}) and 2.33\% for in-air gestures (\textit{One Hand}, \textit{Two Hands}, and \textit{Body Rotation}), which is consistent with previous work\cite{Morar2020Evaluation, Kuhlmann2021}.}

\begin{table}[!htbp]
\footnotesize
\renewcommand{\arraystretch}{1.3}
\caption{\chadded{Inherent error of AngleSizer. The T columns indicate target scales, and the R columns indicate \textbf{Average Value (Standard Deviation)} of measured results. The last row illustrates the \textbf{Average Relative Error}(ARE) of each gesture.}}
\label{Sighted Measure Accuracy}
\centering
\resizebox{\textwidth}{!}{
\begin{threeparttable}
\begin{tabular}{ccccccccccc}
\toprule
&\multicolumn{2}{c}{\textbf{One Finger\tnote{*}}} & \multicolumn{2}{c}{\textbf{Two Fingers\tnote{*}}} & \multicolumn{2}{c}{\textbf{One Hand\tnote{*}}} & \multicolumn{2}{c}{\textbf{Two Hands\tnote{*}}} & \multicolumn{2}{c}{\textbf{Body Rotation\tnote{\dag}}} \\
&\textbf{T} & \textbf{R} & \textbf{T} & \textbf{R} & \textbf{T} & \textbf{R} & \textbf{T} & \textbf{R} & \textbf{T} & \textbf{R} \\
\midrule
&2.0              & 1.99(0.057)               & 2.0                & 1.98(0.063)              & 20              & 21.1(1.595)             & 20                 & 20.0(0.000)                 & 60             & 64.8(2.781)           \\
&4.0              & 3.97(0.048)               & 4.0                & 3.96(0.097)              & 40                & 39.0(1.886)             & 40                 & 40.0(0.000)                 & 120            & 118.4(6.041)           \\
&6.0              & 5.94(0.084)               & 6.0                & 5.98(0.063)              & 60              & 59.4(1.075)             & 60               & 60.1(0.738)             & 180            & 183.7(6.056)           \\
&8.0              & 7.95(0.071)            & 8.0                & 7.98(0.092)              & 80              & 78.4(2.757)             & 80               & 80.1(0.876)             & 240              & 242.0(5.395)           \\
&10.0              & 9.92(0.162)               & 10.0                & 9.89(0.145)              & 100              & 98.2(1.874)             & 100              & 101.1(1.101)             & 300            & 303.1(5.527)           \\
&12.0             & 11.95(0.108)               & 12.0               & 11.86(0.151)              & 120             & 120.3(1.829)             & 120              & 121.3(1.494)             & 360            & 362.5(3.689)    \\   
\hline
\textbf{ARE} && {\textbf{1.13\%}} & &{\textbf{1.31\%}} & &{\textbf{3.15\%}}& &{\textbf{0.67\%}} & &{\textbf{3.16\%}} \\
\bottomrule
\end{tabular}
\begin{tablenotes}\footnotesize
\item[*] All values are in centimeters (cm).
\item[\dag] All values are in degrees (°).
\end{tablenotes}
\end{threeparttable}
}
\end{table}

\chadded{We recruited 7 sighted participants (4 males, 3 females) aged between 22 and 30 ($M$ = 26.71, $SD$ = 2.91) to evaluate the \textit{operational error} of AngleSizer in real scenarios. Before the experiment, participants were asked to familiarize themselves with the usage of AngleSizer for about 10 minutes. During the experiment, participants were blindfolded and asked to use the free exploration module to measure a series of objects' sizes and rotate by some angles with the five interaction gestures. As shown in Table~\ref{Sighted Measure Operation Accuracy}, the average operational errors of AngleSizer are 4.88\% for on-screen gestures and 3.83\% for in-air gestures. 
The higher operational error stems from variations in participant behavior (e.g., the inability to maintain the smartphone in a strictly parallel orientation to the ground during body rotations and observable hand jitter at the completion of each gesture). However, these deviations remained within the acceptable tolerances of the system.}

\begin{table}[h]
\footnotesize
\caption{\chadded{Operational error of AngleSizer. The T columns indicate accurate scales of target objects, and the R columns indicate \textbf{Average Value (Standard Deviation)} of measured results. The last row illustrates the Average Relative Error (ARE) of each gesture.}}
\label{Sighted Measure Operation Accuracy}
\centering
\resizebox{\textwidth}{!}{
\begin{threeparttable}
\begin{tabular}{ccccccccccc}
\toprule
&\multicolumn{2}{c}{\textbf{One Finger\tnote{*}}} & \multicolumn{2}{c}{\textbf{Two Fingers\tnote{*}}} & \multicolumn{2}{c}{\textbf{One Hand\tnote{*}}} & \multicolumn{2}{c}{\textbf{Two Hands\tnote{*}}} & \multicolumn{2}{c}{\textbf{Body Rotation\tnote{\dag}}} \\
&\textbf{T} & \textbf{R} & \textbf{T} & \textbf{R} & \textbf{T} & \textbf{R} & \textbf{T} & \textbf{R} & \textbf{T} & \textbf{R} \\
\midrule
&1.8              & 1.89(0.121)               & 1.8                & 1.80(0.183)              & 23              & 22.7(1.254)             & 23                 & 22.7(0.756)                 & 45             & 43.7(3.352)           \\
&3.0              & 3.09(0.157)               & 3.0                & 3.11(0.302)              & 36                & 35.9(1.773)             & 36                 & 36.4(1.272)                 & 60            & 59.9(3.671)           \\
&5.0              & 5.03(0.243)               & 5.0                & 4.83(0.368)              & 60              & 58.4(2.637)             & 60               & 59.9(4.298)             & 90            & 87.9(3.671)           \\
&7.6              & 7.54(0.223)            & 7.6                & 7.43(0.229)              & 78              & 74.6(2.820)             & 78               & 77.6(2.370)             & 135              & 133.9(5.728)           \\
&9.3              & 9.51(0.121)               & 9.3                & 9.36(0.310)              & 109              & 109.6(3.409)             & 109              & 106.4(4.315)             & 180            &179.7(4.461)   \\
\hline
\textbf{ARE} && {\textbf{3.93\%}} & &{\textbf{5.83\%}} & &{\textbf{3.77\%}}& &{\textbf{3.45\%}} & &{\textbf{4.27\%}} \\
\bottomrule
\end{tabular}
\begin{tablenotes}\footnotesize
\item[*] All values are in centimeters (cm).
\item[\dag] All values are in degrees (°).
\end{tablenotes}
\end{threeparttable}
}
\end{table}

\section{Study 2: Evaluation of AngleSizer} \label{Cap:User study}
We conducted a user study to evaluate the efficacy of AngleSizer in helping BVI users acquire and enhance spatial scale perception. In the evaluation, we first defined 20 tasks corresponding to 5 interaction gestures to assess the perception ability of the participants. Then, we conducted a within-subjects study to compare the spatial perception accuracy before and after using AngleSizer and to evaluate the learning effect of using different gestures. 
We omitted a baseline control group since the pre-research user performance served as the initial state without AngleSizer intervention.


\subsection{Participants}
We recruited 11 participants (7 males, 4 females) from the local community, with an average age of 32.182 (SD = 7.791). Among them, five were totally blind, three had blindness with light perception, and three had low vision (Table~\ref{tab4:Demographic 14}). 
Each participant was compensated \$150 upon completion of the study.

\begin{table*}  [h]
\small
\caption{Self-reported demographic information of 11 participants of the user study.}  \label{tab4:Demographic 14}
 \setlength{\tabcolsep}{6pt}{
  \begin{tabular}{cccllcc}
    \toprule
    \textbf{No.} & \textbf{Age} & \textbf{Gender} & \textbf{\makecell{Visual Condition}} & \textbf{\makecell{Phone Model}} & \textbf{\makecell{Educational\\Background}} & \textbf{\makecell{Residential\\Status}}\\
    \midrule
    \ P1 & 24 & M & blind & Honor Nova 6 & bachelor's & with roommates \\
    \ P2 & 27 & M & blind with light perception & Mi 10 Pro & bachelor's & with roommates \\
    \ P3 & 28 & M & blind & Huawei Mate 40 Pro & bachelor's & with wife \\
    \ P4 & 38 & F & low vision & Mi 6 & bachelor's & with husband \\
    \ P5 & 50 & M & low vision & Huawei P9 & bachelor's & with wife \\
    \ P6 & 43 & F & blind & Redmi Note 9 & bachelor's & alone  \\
    \ P7 & 25 & F & low vision & Mi 11 & bachelor's & with family \\
    \ P8 & 32 & M & blind & Redmi Note 8 & bachelor's & with girlfriend \\
    \ P9 & 28 & M &blind with light perception & Mi 10s & bachelor's & with colleagues \\
    \ P10 & 31 & M & blind with light perception & Redmi K30 & bachelor's & with colleagues \\
    \ P11 & 28 & F & blind & Mi 6 & bachelor's & with colleagues \\
  \bottomrule
\end{tabular}}
\end{table*}

\subsection{Task Design}
\label{sec:taskdesign}
Since the potential values for size and angle measurements are continuous, it is impractical to exhaustively assess the spatial perception ability of participants. 
For instance, assessing size perception necessitates measuring the proficiency of users at discerning variations across a potentially infinite range, such as 1cm, 2cm, and so forth. 
To address this challenge, we selected a representative set of values for the assessment of spatial perception ability. 

According to the preliminary DTD survey in Section~\ref{Cap:Preliminary research}, participants reported that different values of the measurements did not affect the perceived learning difficulty. 
Therefore, we empirically chose four sizes commonly adopted in daily lives for each of the five gestures as the learning tasks in the evaluation (Table~\ref{tab5:Learning Task}). 
We use the relative error (obtained from the Assessment module) of these selected tasks to indicate the overall ability to perceive size and angles for each participant. 
During the study, we encoded the selected 20 tasks to the Assessment module and required the participants to assess their abilities every day (See Section~\ref{sec:procedure}). 

\begin{table}  [h]
  \caption{Learning tasks of different gestures.}  \label{tab5:Learning Task}
 \setlength{\tabcolsep}{3.8mm}{
  \begin{tabular}{lllll}
    \toprule
    \textbf{Gestures} & \textbf{Task} &    &    &    \\
    \midrule
    \textbf{One Finger} & 1cm & 2cm & 4cm & 8cm \\
    \textbf{Two Fingers} & 3cm & 5cm & 6cm & 7cm \\
    \textbf{One Hand} & 25cm & 35cm & 70cm & 100cm \\
    \textbf{Two Hands} & 45cm & 65cm & 85cm & 100cm \\
    \textbf{Body Rotation} & 30° & 45° & 60° & 120° \\
  \bottomrule
\end{tabular}}
\end{table}

\subsection{Procedure}\label{sec:procedure}
\begin{figure*}[ht] 
  \centering
  \includegraphics[width=\textwidth]{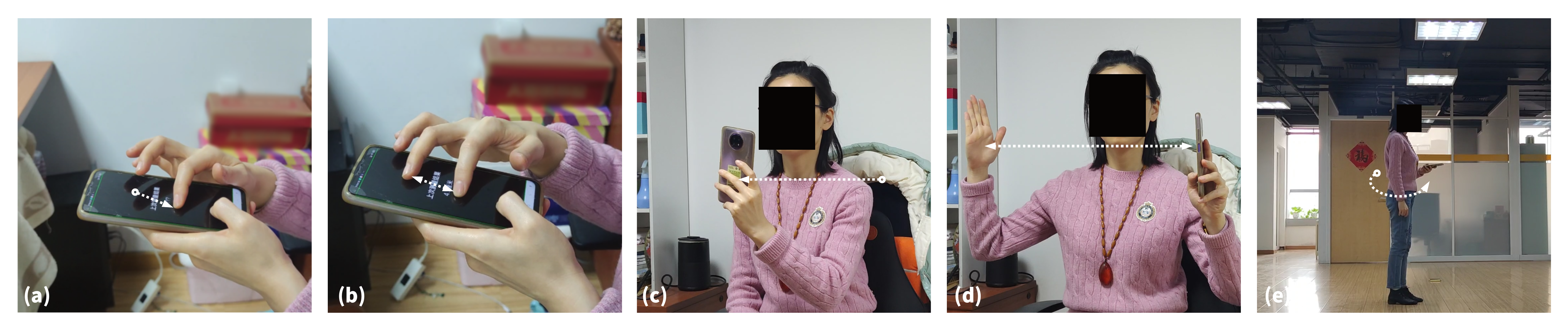}
  \caption{A participant is using AngleSizer. (a) Using the one-finger gesture, with the right hand's index finger moving on the screen. (b) Using the two-finger gesture, to measure the distance between the right hand's index and middle fingers on the screen. (c) Using the one-hand gesture, to measure the distance moved by the left hand holding the phone. (d) Using the two-hand gesture. (e) Using the body-rotation gesture, holding the phone parallel to the ground and rotating by an angle.} \label{Fig: Experiment}
  \Description{Process flow of the system at each frame, including getting raw data, checking validity, computing, checking stability, and giving feedback.}
\end{figure*}

The study was conducted over a period of ten consecutive days.
On the first day, we installed AngleSizer face-to-face on the personal smartphone of each participant. After the introduction of the basic usage of AngleSizer, participants were asked to use AngleSizer for 20 minutes to familiarize themselves with the interaction (Figure \ref{Fig: Experiment}). Then each participant was required to measure their initial spatial perception ability using the \textit{Ability Assessment} module.

Afterward, participants were required to use AngleSizer in their daily lives for 10 days. They were required to learn and explore all of the five gestures every day. They had the flexibility to freely switch among three modules and were encouraged to discover more space dimensions beyond preset tasks while exploring their surroundings. Specifically, participants were required to use the \textit{Ability Assessment} module at the end of each day to measure their spatial perception ability. During the study, we recorded each interaction, along with the timestamp, task, and assessment results from participants. Each time the participant opened AngleSizer, the log files were uploaded automatically to the server through the network.

Finally, self-assessment questionnaires and interviews were carried out to gather subjective feedback.

\section{Results}
During the evaluation study, we collected 10,849 instances of usage (each instance indicated one task either in the \textit{Guided Learning}, \textit{Free Exploration}, or \textit{Capability Assessment} module), cumulatively amounting to 6,482.317 minutes in total. On average, each participant used AngleSizer for 98.627 times a day (84.491 times in the \textit{Assessment} module, 9.773 times in the \textit{Exploration} module, and 3.363 times in the \textit{Learning} module), which was approximately one hour of daily interaction with AngleSizer.

\subsection{Spatial Perception Ability Improvement}
As mentioned in Section~\ref{sec:taskdesign}, we measured the spatial perception ability of the participants using the average relative error of selected tasks, which was calculated as \(\frac{\|operation\ result - task\ value\|}{\|task\ value\|}
\).

\begin{table*}[h]  
\footnotesize
   \renewcommand\arraystretch{1.3}
    \setlength{\tabcolsep}{5.5mm}{
  \caption{Mean (standard deviation) of relative error across different gestures. The results of paired t-tests demonstrated that all observed differences were statistically significant.}\label{tab:results}
  \begin{tabular}{ccccc}
    \toprule

\multirow{2}{*}{\textbf{Gestures}} & \multicolumn{2}{c}{\textbf{Relative Error}} & \multirow{2}{*}{\textbf{\thead{$\mathbf{t_{10}}$/$\mathbf{t_{54}}$}}} &  \multirow{2}{*}{\textbf{p-value}}\\
                   &Pre-study        & Post-study       &                   &  \\
    
    \midrule
    One Finger & 0.213 (0.085) & 0.073 (0.038)&-5.306 & <.001\\ 
    Two Fingers & 0.184 (0.069) & 0.060 (0.025)&-5.561&<.001\\
    One Hand & 0.317 (0.091) & 0.174 (0.099)&-3.087&<.01\\
    Two Hands & 0.177 (0.074) & 0.065 (0.029)&-4.714&<.001\\
    Body Rotation & 0.280 (0.100) & 0.115 (0.059)&-4.360&<.001 \\ \midrule
   Overall & 0.234 (0.098) & 0.097 (0.070)&-9.585&<.001\\
  \bottomrule
\end{tabular}}
\end{table*}

Table~\ref{tab:results} illustrates the mean and standard deviation of relative errors before and after the evaluation study. Paired t-test results revealed a statistically significant reduction in relative error for each gesture, underscoring the efficacy of AngleSizer. On average, there was a substantial decrease in relative error by 58.547\%, dropping from 0.234 to 0.097 following a 10-day usage of AngleSizer. Among the five gestures, the \textit{Two Fingers} gesture exhibited the most remarkable reduction of 67.391\%, and the \textit{One Hand} gesture demonstrated the least reduction at 45.110\%. This differential impact across gesture types highlights the varying levels of effectiveness of AngleSizer in enhancing spatial perception ability through different modes of interaction.

\subsection{Learning Effect}

\begin{figure} 
  \centering
  \includegraphics[width=0.85\textwidth]{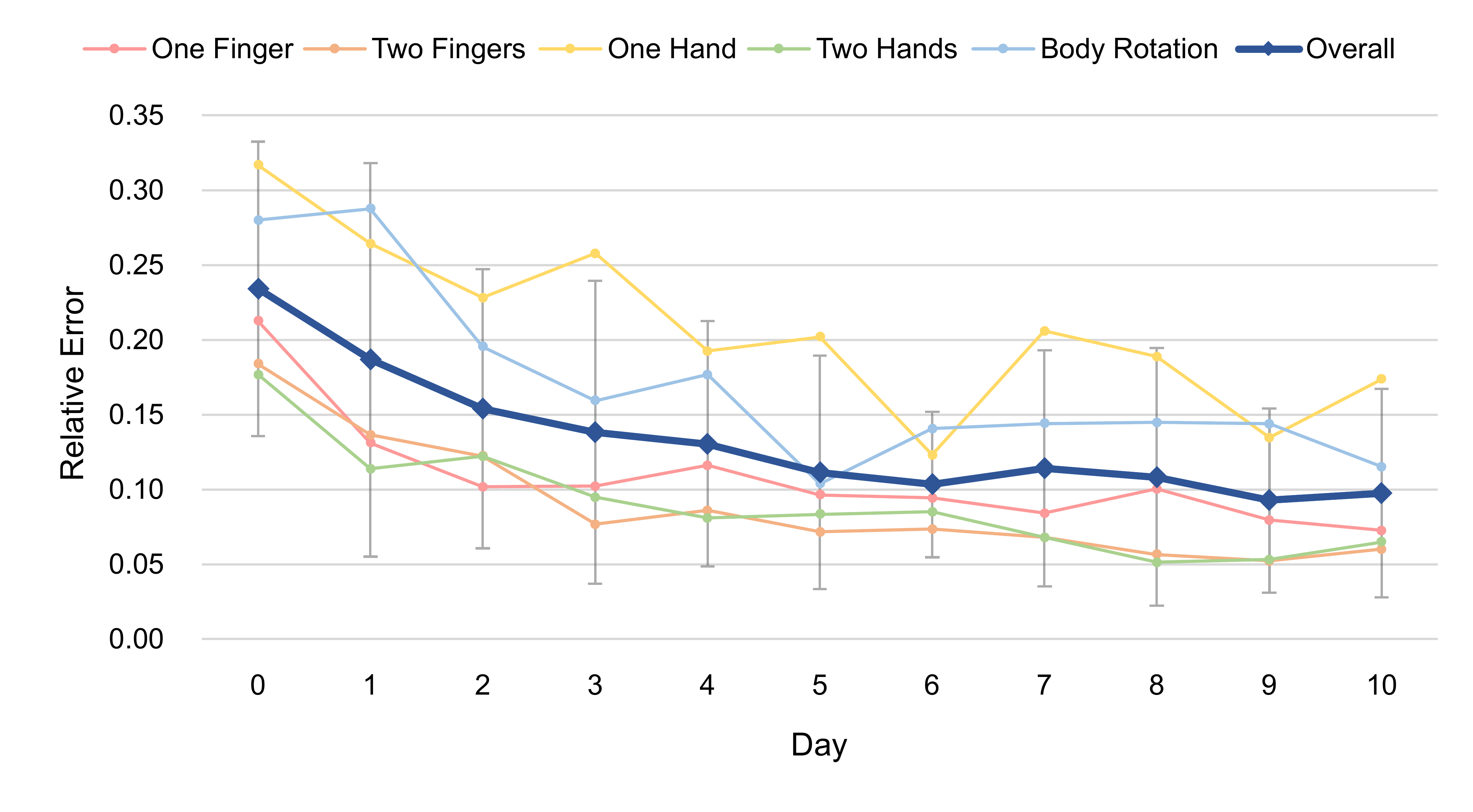}
  \caption{The average relative error on each day. Day 0 represents the initial pre-study ability assessment results. The error bars indicate the standard deviations.}        \label{Fig:errorvstime}
  \Description{The relative error on each day. Day 0 represents the initial pre-study ability assessment. The error bars indicate the standard deviations.}
\end{figure}

To investigate the learning effect using AngleSizer, we conducted a repeated measures ANOVA (RM-ANOVA) with two factors: \textit{Time} and \textit{Gesture}. RM-ANOVA indicated that there was a significant main effect of \textit{Time} on relative error ($F_{10,100}=21.601, p < .001$), and a significant main effect of \textit{Gesture} on relative error ($F_{4, 40} = 27.911, p < .001$).
No significant interaction was found between \textit{Time} and \textit{Gesture}. Figure~\ref{Fig:errorvstime} illustrates the temporal evolution of participants' relative errors throughout the study. Detailed learning curves of each gesture are depicted in the appendix, as shown in Figure~\ref{Fig: Each Gesture Curve}.


The post-hoc pairwise comparison of \textit{Gesture} factor led to the categorization of gestures into two distinct groups:
\begin{itemize}
    \item The first group, which we referred to as \textbf{``gestures with reference''}, encompasses \textit{One Finger}, \textit{Two Fingers}, and \textit{Two Hands}. Gestures within this group were characterized by their ease of control and execution, attributable to the potential for reference through either a touchscreen interface (in the case of \textit{One Finger}) or proprioceptive feedback (applicable to \textit{Two Hands} and \textit{Two Fingers}).     
    \item The second group, labeled as \textbf{``gestures without reference''}, consists of \textit{One Hand} and \textit{Body Rotation}. The defining characteristic of this group is the reliance on one-dimensional movement executed without any external reference points. This lack of reference poses a greater challenge in accurately executing these gestures, as they depend solely on the user’s ability to gauge and control movement in a single dimension.
\end{itemize}

Following this classification, we proceeded to segregate the data into two distinct groups corresponding to the gesture categories. For each group, we conducted an RM-ANOVA focusing solely on the factor \textit{Time}.

In the group of gestures with reference, the analysis revealed a significant main effect of \textit{Time} on relative error (\(F_{10,100}=24.799, p<.001\)). A subsequent post-hoc pairwise comparison indicated the learning process within this group. It was observed that up until day 4, there was a notable and consistent reduction in relative error. However, after day 4, the relative error did not exhibit any significant changes. This pattern suggests that participants were able to achieve a high level of proficiency in the corresponding tasks within approximately 4 days, subsequently maintaining a low error rate in the days that followed.

In contrast, the gestures without reference also exhibited a significant main effect (\(F_{10,100}=6.962, p<.001\)), but with a distinct learning curve. The post-hoc pairwise comparisons pointed out that only on days 9 and 10 did the relative error significantly drop below the initial levels. This implies that participants required a substantially longer duration, approximately 9 days, to adeptly manipulate the gestures in this more challenging group.

These findings underscore the differential learning trajectories and adaptation rates associated with the two gesture groups. The reference gesture group, aided perhaps by proprioceptive feedback, allowed for a quicker learning and mastery process. In contrast, the no reference gesture group, lacking such intuitive physical references, necessitated a longer period for participants to acclimate and achieve proficiency. Participant feedback corroborated these results, with many (e.g., P3, P6, P8) reporting greater ease and quicker learning with gestures with reference compared to the ones without reference.

\subsection{Subjective Feedback} \label{Cap: 5.4 Subjective Feed}

We adapted the NASA scale to evaluate the user experience of AngleSizer, as outlined in Table \ref{tab7:Participants' reviews}, emphasizing users' positive feedback regarding its usability and efficiency. Figure \ref{Fig: Self Evaluation}, derived from the \textit{Rosenberg Self-Esteem Scale (RSE)}, depicted participants' self-assessment before and after their learning process. Through these questionnaires results and interviews, we conducted comprehensive summaries and gleaned valuable insights into their learning journey.). Timely feedback and corrective guidance is the key to training, which greatly improves the learning efficiency of users.

\subsubsection{{\textbf{Factors contributing to learning progress.}}}
Timely feedback is identified as crucial for training, with users emphasizing AngleSizer's real-time feedback on movement sizes or angles, coupled with flexible gesture switching, leading to a richer and more accurate perception, significantly improving learning efficiency. Participants also shared effective learning methods for different tasks, such as \textit{``experiencing the feeling of arm soreness''} (P4), \textit{``feeling the range of motion of hand swing''} (P8), \textit{``feeling the degree of bending of fingers and arms, and the duration of exercise''} (P1), and \textit{``comprehensively judging combined with surrounding objects and space''} (P5). Moreover, the cooperation of the three modules has narrowed the gap between cognition and practice, and increased their scale reserves (P1, P3-P8, P11). Figure~\ref{Fig: Self Evaluation} illustrates the subjective rating differences before and after the study.

\subsubsection{\textbf{Insights from gesture-based learning experiences.}}
We also received feedback about Gesture-based learning experiences. Finger-based gestures, reliant on finger joint angles and screen edges as references, are noted to aid comprehension significantly (P8, P11). The {\itshape Two Fingers} gesture offers mutual references, facilitating easier mastery compared to the {\itshape One Finger} gesture (P3). Similar distinctions were observed between the \textit{One Hand} and \textit{Two Hand} gestures, utilizing various body parts such as eyes, neck, chest, and legs as additional points of reference (P8, P9). Muscle soreness in the arms contributed significantly to the learning process (P2). Exploring angle learning through body rotation extends previous limited plane perception and enhances the body's clarity in perceiving turning and the original position (P5). Furthermore, participants expressed heightened expectations for system capabilities. For instance, P2 suggested enhanced precision in feedback mechanisms, proposing auditory cues like changing frequencies alongside vibrations for more accurate corrections near the target.

\subsubsection{\textbf{Preference for learning modules.}}


The three teaching modes are adapted to different learning needs. Through interviews in the experiment, we learned that when the user has no knowledge of a certain size at all (e.g., P11-45cm, P4-30cm, P9-120degree), he/she tends to use the \textit{Guided Learning} module, which helps him establish an initial cognitive reference. When the user already has a rough sense of the target size(e.g., P1-7cm, P6-3cm, P8-60 degree), but not accurate, the user tends to use the \textit{Ability Assessment} module, which can help him judge the difference between his actual operation and the real size. When users are curious about their surroundings, they tend to use \textit{Free Exploration} modules. On the whole, \textit{Exploration} and \textit{Assessment} are preferred by users. P8 says that the Free exploration module can accumulate the ``knowledge base'' in the exploration of objects in daily life; the process of Assessment is more like a game, full of tension and stimulation, and you can feel achievement when the task is done correctly.

\begin{table*}
\footnotesize
   \renewcommand\arraystretch{1.3}
    \setlength{\tabcolsep}{3.5mm}{
  \caption{Participants' reviews on AngleSizer. 1=Disagree strongly,
5=Agree strongly}  \label{tab7:Participants' reviews}
  \begin{tabular}{lcc}
    \toprule
    \textbf{Opinions} & \textbf{\thead{Average Mark\\ (Max is 5)}} & \textbf{\thead{Standard \\ Deviation}}\\
    \midrule
    The system is helpful for me to know sizes and angles better based on my own parts. & 4.727 & 0.467 \\ 
    It's easy to understand the interaction gestures of AngleSizer. & 4.636 & 0.505\\
    It's easy to understand AngleSizer's modules and plans of teaching. & 4.636 & 0.505\\
    The system's teaching improves my understanding of precise sizes and angles. & 4.455 & 0.522\\
    The system's teaching improves my overall understanding of spatial scales. & 4.545 & 0.522\\
    Real-time feedback improves the efficiency of my studying and exploring sizes and angles. & 4.545 & 0.522\\
    I think it's meaningful to use cell phones to learn about sizes and angles. & 4.455 & 0.522 \\
  \bottomrule
\end{tabular}}
\end{table*}

\subsubsection{{\textbf{Positive impact on participants' lives.}}} \label{Positive impact on participants' lives.}

\begin{figure}[h] 
    \centering
    \includegraphics[width=0.9\textwidth]{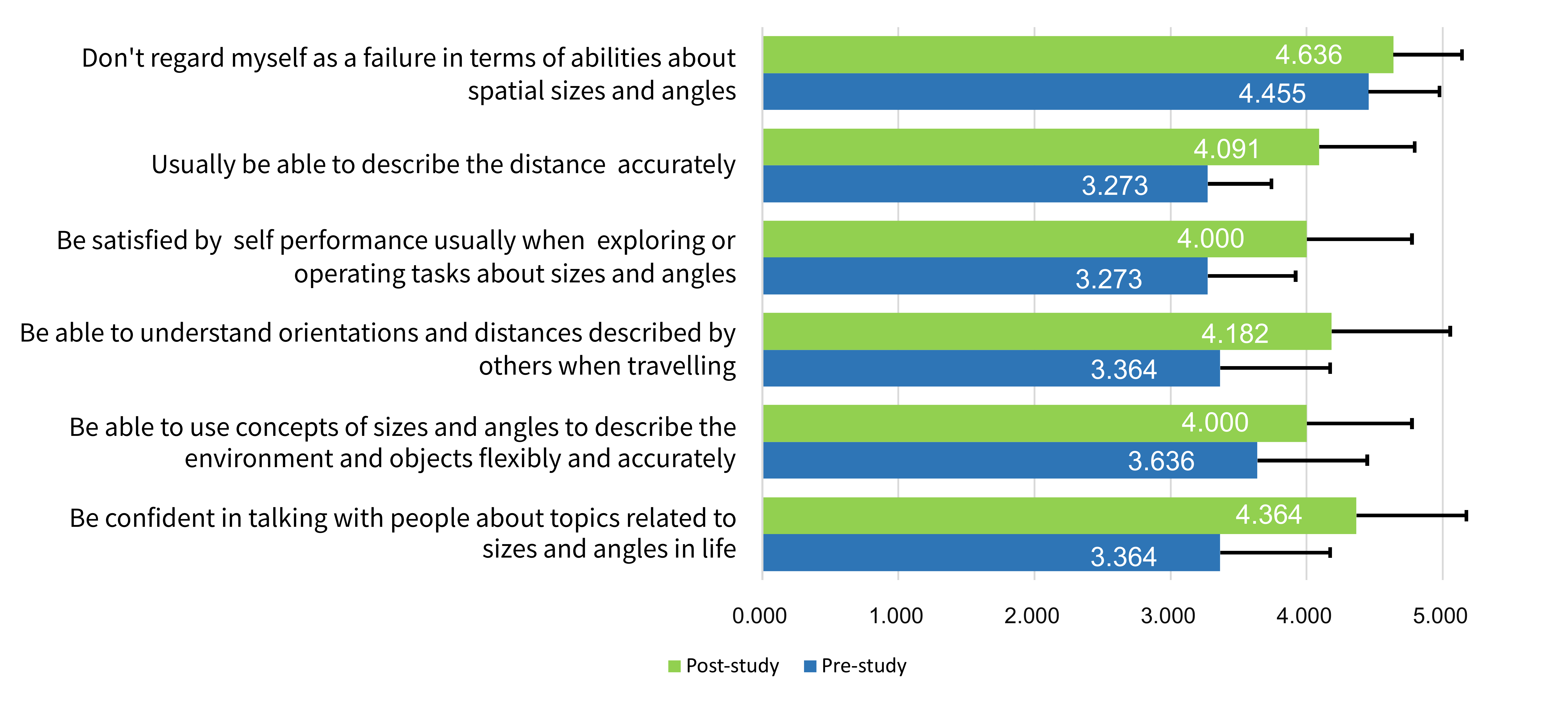}
    \caption{ Bar plot of 11 users' subjective ratings before and after using AngleSizer on a 5-point Likert scale. The error bars indicate the standard deviations.}
    \label{Fig: Self Evaluation}
\end{figure}

The self-ratings changes in Figure \ref{Fig: Self Evaluation} echo these positive shifts. Participants also positively acknowledge the meaning of learning sizes and angles (see Table \ref{tab7:Participants' reviews}). They all mentioned the newfound importance of learning scales, noting a marked improvement in understanding scale information. After learning, especially using the free exploration module, their understanding of scale information became accurate, as P10 said, {\itshape``Compared to my imagination, the real 1 cm is much smaller, and 85 cm is bigger. Many cognitive biases have been reduced.''}. Their information exchange process is much smoother, as P6 mentioned, {\itshape ``In the dance class, I can accurately understand the teacher's movement guidance on how to turn the body makes my learning experience smoother!''} Most importantly, the improved spatial perception ability helped them solve problems related to size in their daily life, as P9 said, {\itshape``I know how wide the music equipment is in reality trough the parameters when purchasing it online.''}

\section{Discussion}

\subsection{Effect of Initial Spatial Perception Ability}

\begin{figure}[h] 
  \centering
  \includegraphics[width=0.6\textwidth]{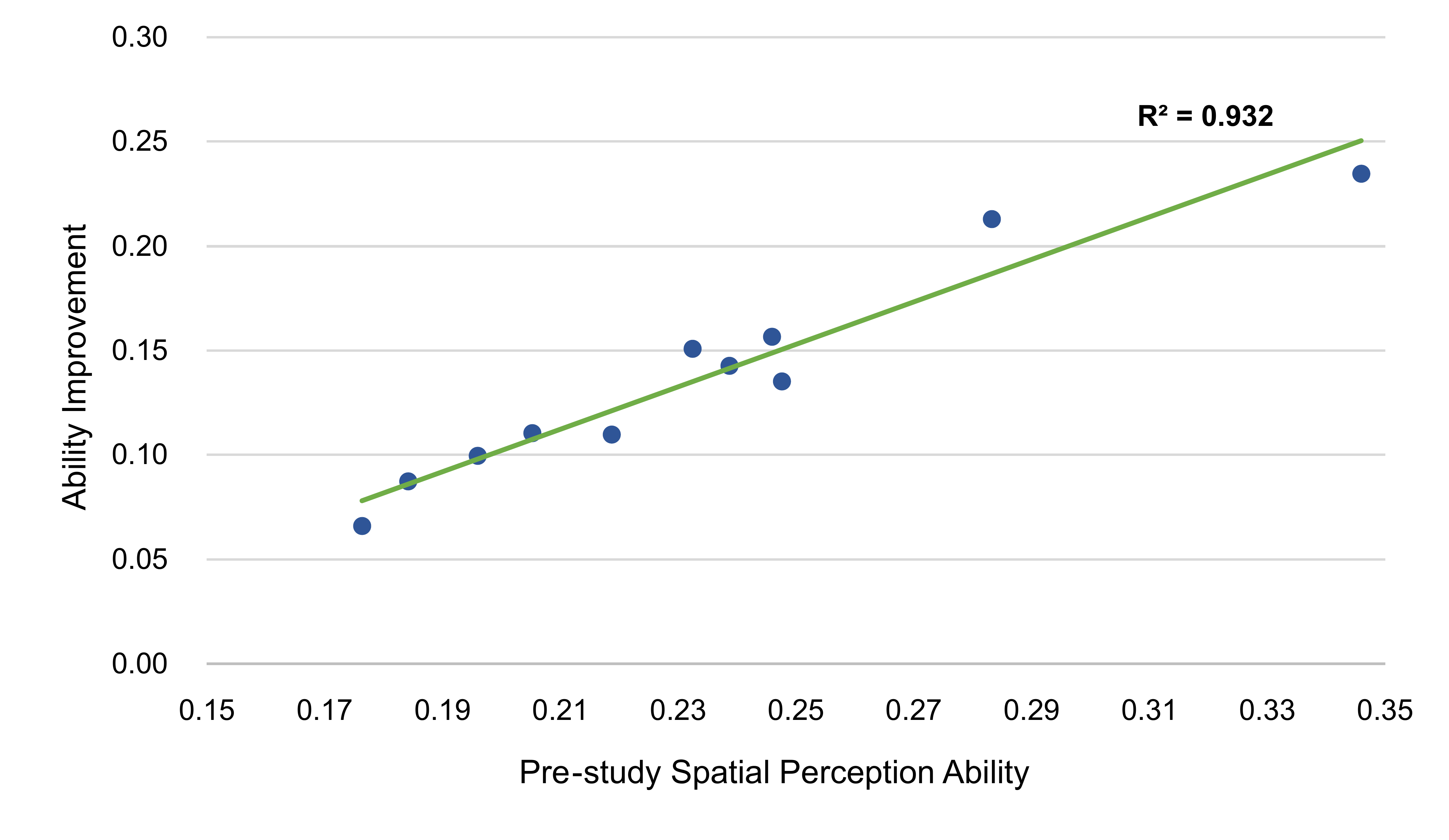}
  \caption{The relationship between participants' pre-study spatial perception abilities and their improvement during the study 2.}        \label{Fig: Improvement Scatter}
  \Description{The relationship between participants' pre-study spatial perception abilities and their improvement during the study 2.}
\end{figure}

Given the inherent variability in spatial perception abilities among different participants, we aimed to explore the extent to which a user's initial ability influences their performance enhancement with AngleSizer.
Therefore, we calculated the difference in relative error before and after the evaluation study and adopted the difference as an indicator of the ability improvement using AngleSizer. We computed the differences for all 11 participants and graphically represented them in Figure~\ref{Fig: Improvement Scatter} alongside their pre-study initial abilities. 

Notably, Figure~\ref{Fig: Improvement Scatter} illustrates a robust linear positive correlation between performance improvement and initial touch accuracy. We conducted a linear regression and depicted the fitted line in the figure,  indicating a high degree of correlation (\(R^2 = 0.932\)). ANOVA results also indicated that the improvement was significantly influenced by initial ability (\(F_{1,9} = 123.594, p<.0001\)).  Consequently, users with lower initial spatial perception abilities tend to experience more pronounced improvements upon using AngleSizer.


\subsection{\chadded{Comparison with Normally Sighted People}}

\chadded{We recruited 7 participants with normal vision (2 males and 5 females) aged between 19 and 30 ($M$ = 25.86, $SD$ = 4.16) to complete the same measurement tasks in the evaluation study (Section~\ref{Cap:User study}) using their bodies. Each task was conducted under two conditions: eyes-opened and blindfolded. Table~\ref{tab7:comparison} illustrates the relative error of sighted users along with the relative error of BVI users before and after using AngleSizer (\textit{Day 0} and \textit{Day 10} in the evaluation study, Section~\ref{Cap:User study}). These results shed light on the typical performance of sighted individuals in tasks involving size and angle estimation. As anticipated, sighted users demonstrated lower errors when visual feedback was engaged.}


\chadded{BVI people lack opportunities for visual feedback, and there is no way to refine this system further, thereby hindering the full development of mobile skills reliant on vision. However, AngleSizer training enhances the proprioceptive system of the blind, with non-visual perception taking precedence in building spatial perception ability. This compensates for the limitations imposed by visual impairments. AngleSizer can help the BVI have equal or even better performances compared with normally sighted people, bringing benefits to sharing information about scales when chatting, as well as activities like dancing in their daily lives (Section~\ref{Positive impact on participants' lives.}).}

\begin{table}[]
\caption{\chadded{Mean (standard deviation) of relative error of BVI users and sighted users.}}\label{tab7:comparison}
\begin{threeparttable}

\begin{tabular}{cccccc}
\toprule
& \textbf{One Finger} & \textbf{Two Fingers} & \textbf{One Hand} & \textbf{Two Hands} & \textbf{Body Rotation} \\
\midrule
\textbf{BVI - Day 0} & 0.213 (0.085) & 0.184 (0.069) & 0.317 (0.091) & 0.177 (0.074) & 0.280 (0.100) \\
\textbf{BVI - Day 10} & 0.073 (0.038) & 0.060 (0.025) & 0.174 (0.099) & 0.065 (0.029) & 0.115 (0.059) \\
\textbf{Sighted - Blindfolded \tnote{*}} & 0.372 (0.051) & 0.567 (0.062) & 0.218 (0.044) & 0.151 (0.043) & 0.177 (0.131) \\
\textbf{Sighted - Eyes-opened} & 0.135 (0.040) & 0.242 (0.093) & 0.213 (0.090) & 0.109 (0.061) & 0.113 (0.072) \\
\bottomrule
\end{tabular}
\begin{tablenotes}\footnotesize
\item[*] Note that this condition is unusual for sighted users.
\end{tablenotes}
\end{threeparttable}
\end{table}


\subsection{Limitations}

AngleSizer is developed to assist the BVI in enhancing their scale perception abilities using current smartphone technologies. We have rigorously tested the system's accuracy and discussed use errors in Section \ref{Cap:System's Accuracy}. Additionally, our preliminary investigations sought to determine the accuracy tolerance required by the BVI for learning scales, as detailed in Section \ref{Cap: Classification of Demands}. While we strive to keep errors within an acceptable range, they are, to some extent, inevitable due to several influencing factors. For instance, when using finger gestures, the large touch area might cause deviations in the coordinates detected by the \textit{TouchEvent} API. The \textit{One Hand} gesture, which relies on SLAM technology, can perform poorly in poorly lit or feature-sparse environments. Moreover, the \textit{Two Hands} and \textit{Body Rotation} gestures work best when the parallel requirement is met, so the beep feedback must play a role. It is worth noting that, our primary contribution is the presentation of a viable design scheme, and we anticipate future implementations that offer higher accuracy and robustness.


Furthermore, there are limitations in our system evaluation. We used \textit{Day 0} as a baseline to assess user ability without stringently controlling for variables like prior knowledge and learning experience, which could impact performance. Future research could provide deeper insights by thoroughly analyzing these factors. Additionally, our evaluation involved only 11 participants, all of whom had advanced educational backgrounds. We plan to conduct a larger, more diverse study in real-world settings to further validate and refine our system.


\subsection{Implications} \label{Cap: 5.4 Subjective Feed}


Our research into the behavioral characteristics and lifestyle patterns of the BVI community enhances our understanding in the field of human-computer interaction (HCI). Looking ahead, future studies could build upon our findings, integrating a wider array of technologies and tools such as wearable devices and white canes to further explore the development of spatial abilities within the BVI community. Additionally, our design concepts and experimental approaches provide valuable insights that could benefit related research fields, including auditory compensation and the training of psychological map construction. We have capitalized on the existing hardware and software capabilities of smartphones to develop interaction gestures, dual-channel feedback, structured teaching processes, and diverse learning modules. Moving forward, these elements can be synergistically combined with mobile sensing technologies \cite{gao2019predicting,gao2023automated} to enhance the efficiency and personalization of how BVI users perceive and interact with their environment.


AngleSizer, as an intelligent teaching assistant, offers portability, universality, and cost-effectiveness, making it a powerful tool for enabling blind individuals to independently learn about sizes and angles. This system has the potential for broader application, such as being incorporated into the curriculum of schools for the blind and extended to more complex subjects like stage performance, geometry, and arithmetic. Such integration would enhance the self-learning capabilities of BVI students. Additionally, it provides support for expanding the user sample to further explore the ability performance for both blind and sighted individuals, paving the way for future comparative studies.

\section{Conclusion} \label{Cap:CONCLUSION}

Spatial perception is crucial for the visually impaired, encompassing understanding large spaces, volume and weight assessment, and discerning details. \chadded{Blindness deprives individuals of opportunities for visual feedback, and refining this system remains a challenge, hindering the full development of mobile skills critical for movement. Additionally, the limited opportunities for physical activity among blind people further constrain proprioceptive development. When comparing the proprioceptive abilities of blind and sighted individuals, visually impaired individuals demonstrate lower accuracy in proprioceptive skills \cite{fiehler2009early,holllns1988spatial}}.
\chadded{ Current digital methods primarily focus on large-scale applications and do not directly address spatial perception training. Meanwhile, traditional physical teaching aids for spatial perception are costly.} 

This paper addresses this gap by concentrating on small and medium spatial scales, introducing AngleSizer as a solution to enhance the spatial scale ability of blind individuals. We conducted preliminary research to identify deficiencies in BVI abilities, drawing inspiration from traditional methods. AngleSizer utilizes smartphones to effectively improve fundamental scale perception, providing timely feedback and precise guidance with ease of operation. Evaluating 11 BVI participants over 10 days with various tasks, we observed positive teaching effects, indicating suitability for individuals with different ability bases. Our findings contribute to the HCI community's understanding of BVI requirements and behaviors, enhancing spatial ability. AngleSizer also positively influences curiosity, self-confidence, spatial perception, and communication abilities, affirming its sustained impact.

\begin{acks}
This work is supported by the Natural Science Foundation of China (Grant No. 62132010 and No. 62302252), Beijing Key Lab of Networked Multimedia, Institute for Artificial Intelligence, Tsinghua University (THUAI), the China Postdoctoral Science Foundation (Grant No. 2023M731949), and Beijing National Research Center for Information Science and Technology (BNRist). 

\end{acks}

\bibliographystyle{ACM-Reference-Format}
\bibliography{1-main-references}

\appendix

\section{Online Questionnaire}
\label{app:questionnaire}
{The following is the questionnaire we used for investigation. \textbf{SO} stands for single-option questions; \textbf{MO} stands for normal multiple-option questions; \textbf{FB} stands for fill-in-blank questions; \textbf{MO-FB} stands for multiple-option questions whose last option allows the recipient to fill in custom answer; \textbf{LS} for 5-point Likert scale questions; \textbf{RK} for ranking questions.}

{\noindent 1. What's your gender? [SO]}

{\indent a) Male; \indent b) Female}

{\noindent 2. What's your age? [FB]}

{\noindent 3. What's your visual condition? [SO]}

{\indent a) Blind; \indent b) Blind with light perception; \indent c) Only able to distinguish some outlines;}

{\indent d) Only able to distinguish some colors and outlines; \indent e) Low-vision}

{\noindent 4. When did you get the visual impairment? [SO]}

{\indent a) Since born; \indent b) When 0 to 3 years old; \indent c) When 4 to 6 years old;}

{\indent d) When 7 to 12 years old; \indent e) When 13 to 18 years old; \indent f) After 18 years old}

{\noindent 5. How long has your visual impairment lasted? [SO]}

{\indent a) For less then 1 year; \indent b) For 1 to 2 years; \indent c) For 3 to 5 years;}

{\indent d) For 6 to 10 years; \indent e) For more than 10 years}

{\noindent 6. Where do you usually live? [SO]}

{\indent a) In a first-tier city; \indent b) In the capital of the province; \indent c) In a prefecture-level city;}

{\indent d) In a county-level city; \indent e) In the rural area; \indent f) Abroad}

{\noindent 7. What's the type of your education? [SO]}

{\indent a) Special education for the BVI; \indent b) Common education; \indent c) Both special and common education}

{\indent d) Home education; \indent e) No education }

{\noindent 8. What's your educational level? [SO]}

{\indent a) Primary or below; \indent b) Junior high; \indent c) Senior high;}

{\indent d) Associate; \indent e) Undergraduate; \indent f) Graduate or above}

{\noindent 9. What is your job? [FB]}

{\noindent 10. What are your hobbies? [FB]}

{\noindent 11. Who are the majority of your friends? [SO]}

{\indent a) BVI people; \indent b) Sighted people; \indent c) People with other disabilities; \indent d) Few friends}

{\noindent 12. Do you recognize yourself as the reference to understand the relationship between yourself and the surrounding objects? [LS]}

{\noindent 13. Do you recognize surrounding objects as references to understand the relationship between yourself and them? [LS]}

{\noindent 14. Do you recognize surrounding objects as references to understand the relationships among them? [LS]}

{\noindent 15. Can you accurately understand the scales of an object or a space described by others? [LS]}

{\noindent 16. Can you describe accurately the scales of an object or a space to others? [LS]}

{\noindent 17. Are you confident when describing the scales? [LS]}

{\noindent 18. Do you often come into problems because of being unable to accurately understand scale-related concepts in daily life? [LS]}

{\noindent 19. Do you have a clear understanding of scale-related concepts? [LS]}

{\noindent 20. Do you intentionally accumulate knowledge about sizes and angles in daily life? [LS]}

{\noindent 21. What contributes to your understanding of sizes and angles? [MO-FB]}

{\indent a) Visual experience in the past; \indent b) Abstract concepts; \indent c) Comparison with objects in daily life;}

{\indent d) Comparison with body parts; \indent e) Correction from sighted people; \indent g) Exploration after error; \indent h) Others}

{\noindent 22. Have you received special training on scale perception? [SO]}

{\indent a) Yes; \indent b) No}

{\noindent 23. What would you do when you want to know a size or angle? [RK]}

{\indent a) Abandon; \indent b) Ask sighted people to measure; \indent c) Compare with body parts;}

{\indent d) Compare with objects nearby; \indent e) Estimation with spatial perception instead of comparison;}

{\indent f) Ruler or other measurement tools}

{\noindent 24. What expressions will be better for you to communicate with others about scales? [SO]}

{\indent a) Specific values, such as 1cm, 10deg; \indent b) Unspecific words, such as a little bit short;}

{\indent c) Comparison expression, such as the same length as the phone}

{\noindent 25. Would you like to use a mobile app to learn about sizes and angles with your body sensation if it exists? [SO]}

{\indent a) Yes; \indent b) No; \indent c) Not sure}

{\noindent 26. In what aspects do you think learning and mastering concepts of sizes and angles can offer convenience? [MO-FB]}

{\indent a) Communication; \indent b) Living independently; \indent c) Spatial Imagination;}

{\indent d) Society integration; \indent e) Information Synchronization with sighted people; \indent f) Others}

{\noindent 27. In what scenarios do you think learning learning and mastering concepts of sizes and angles can help? [FB]}

{\noindent 28. Would you like to participate in our further study such as interviews and user study? [SO]}

{\indent a) Yes; \indent b) No}

\section{The relative error curves of each {\itshape Gesture}}

\begin{figure*} [h]
    \centering
    \begin{minipage}[t]{0.33\linewidth}
        \centering
        \includegraphics[width=\linewidth]{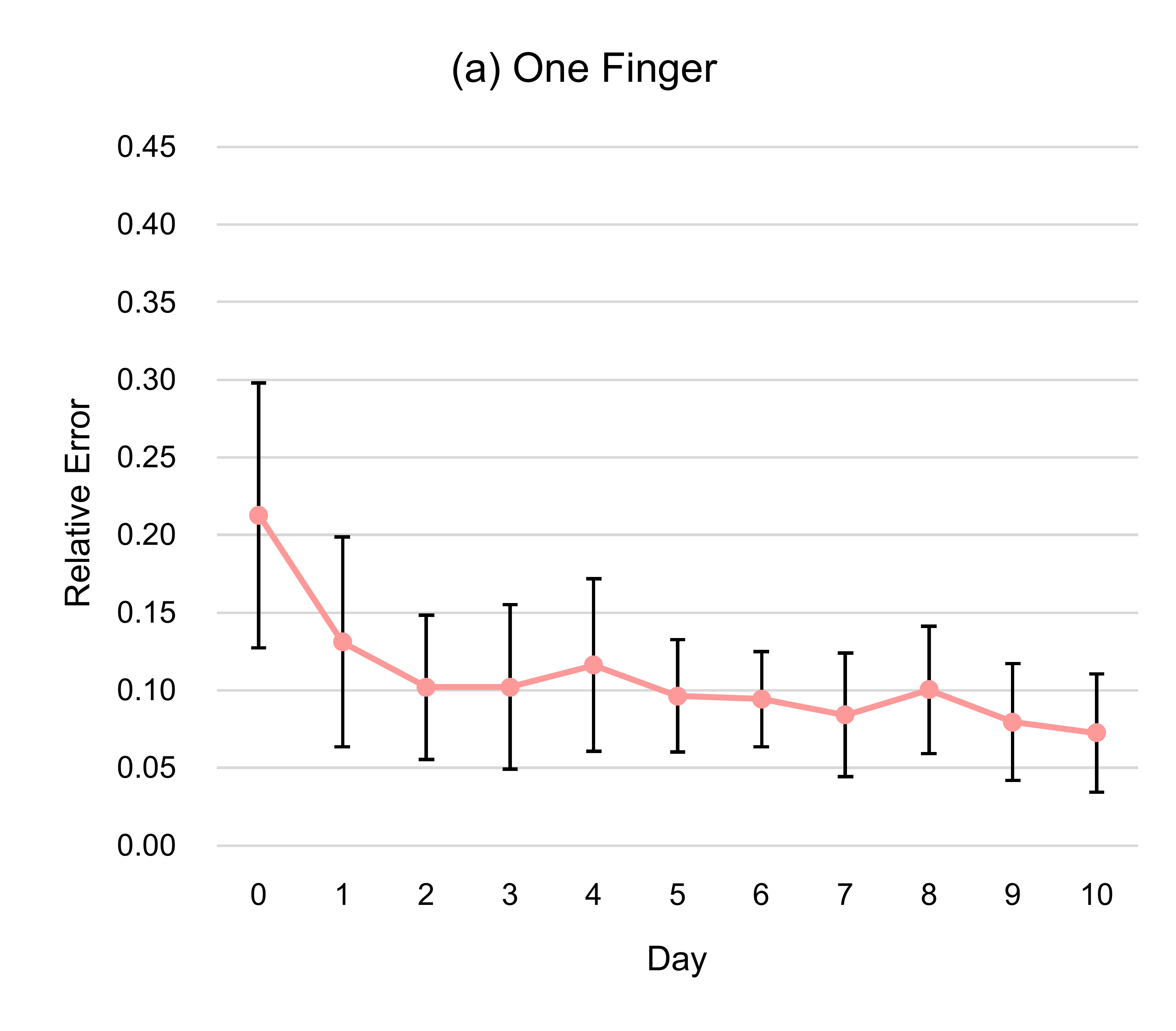}
    \end{minipage}%
    \begin{minipage}[t]{0.33\linewidth}
        \centering
        \includegraphics[width=\linewidth]{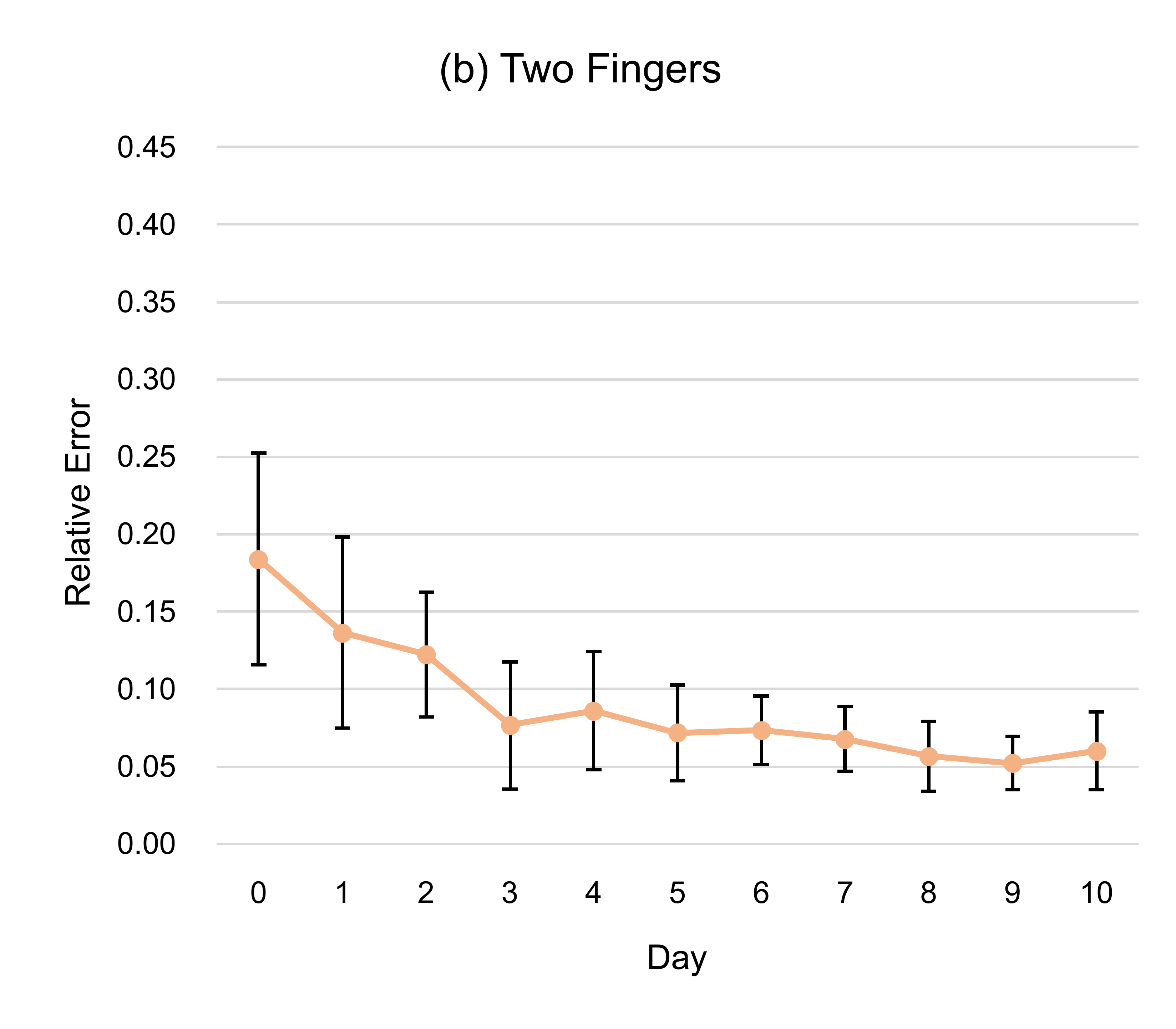}
    \end{minipage}%
    \begin{minipage}[t]{0.33\linewidth}
        \centering
        \includegraphics[width=\linewidth]{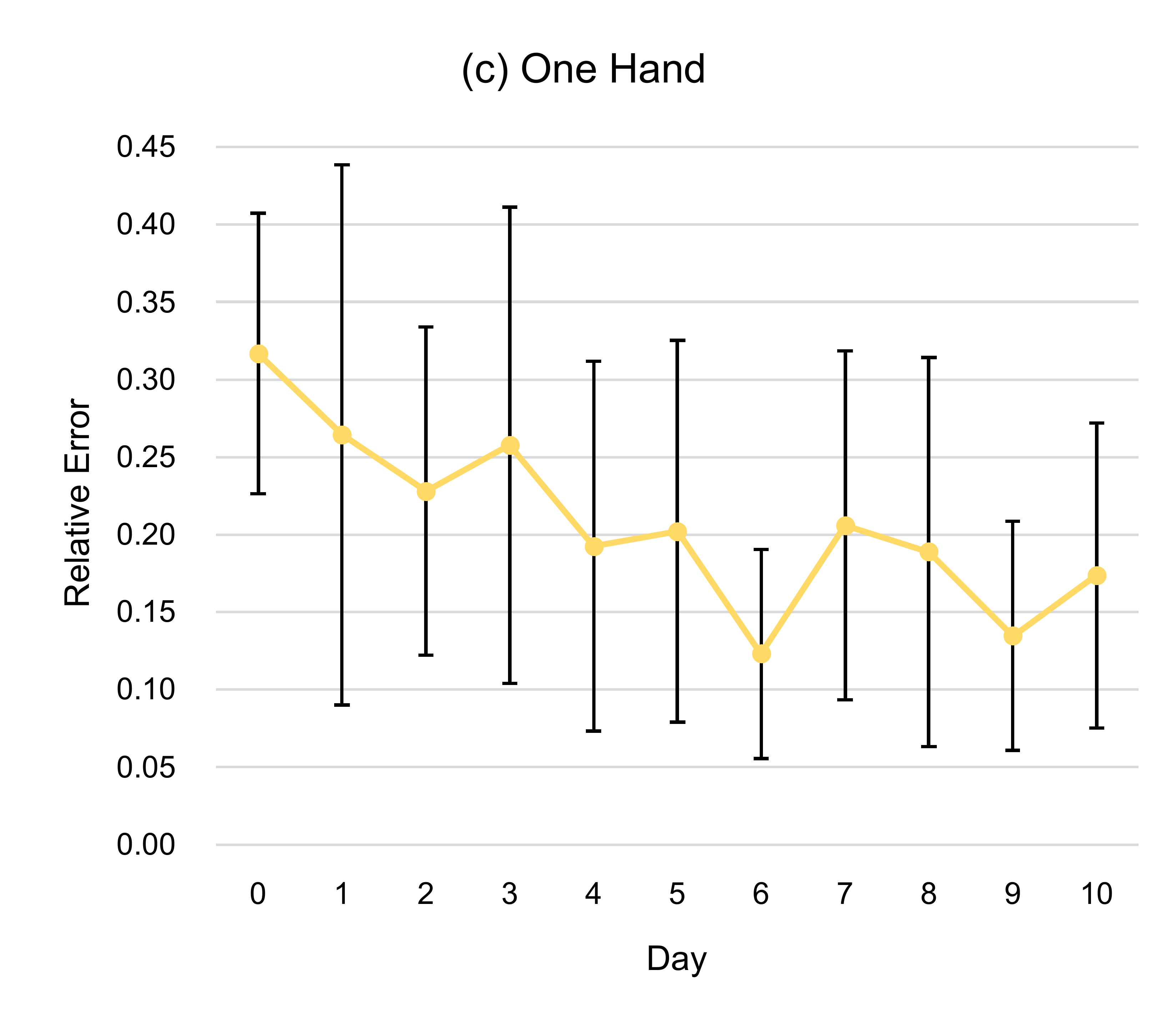}
    \end{minipage}%
    \\
    \begin{minipage}[t]{0.33\linewidth}
        \centering
        \includegraphics[width=\linewidth]{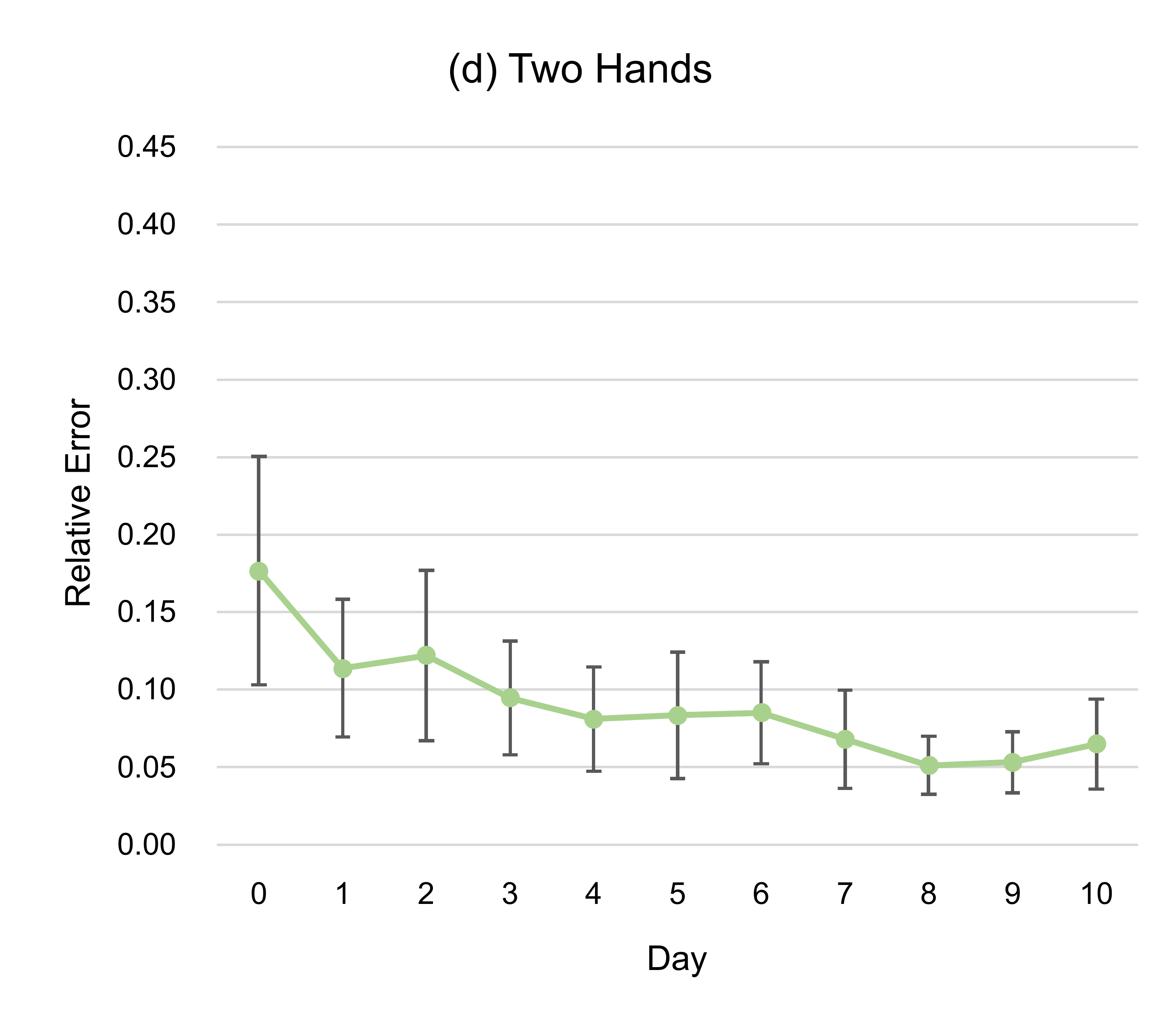}
    \end{minipage}%
    \begin{minipage}[t]{0.33\linewidth}
        \centering
        \includegraphics[width=\linewidth]{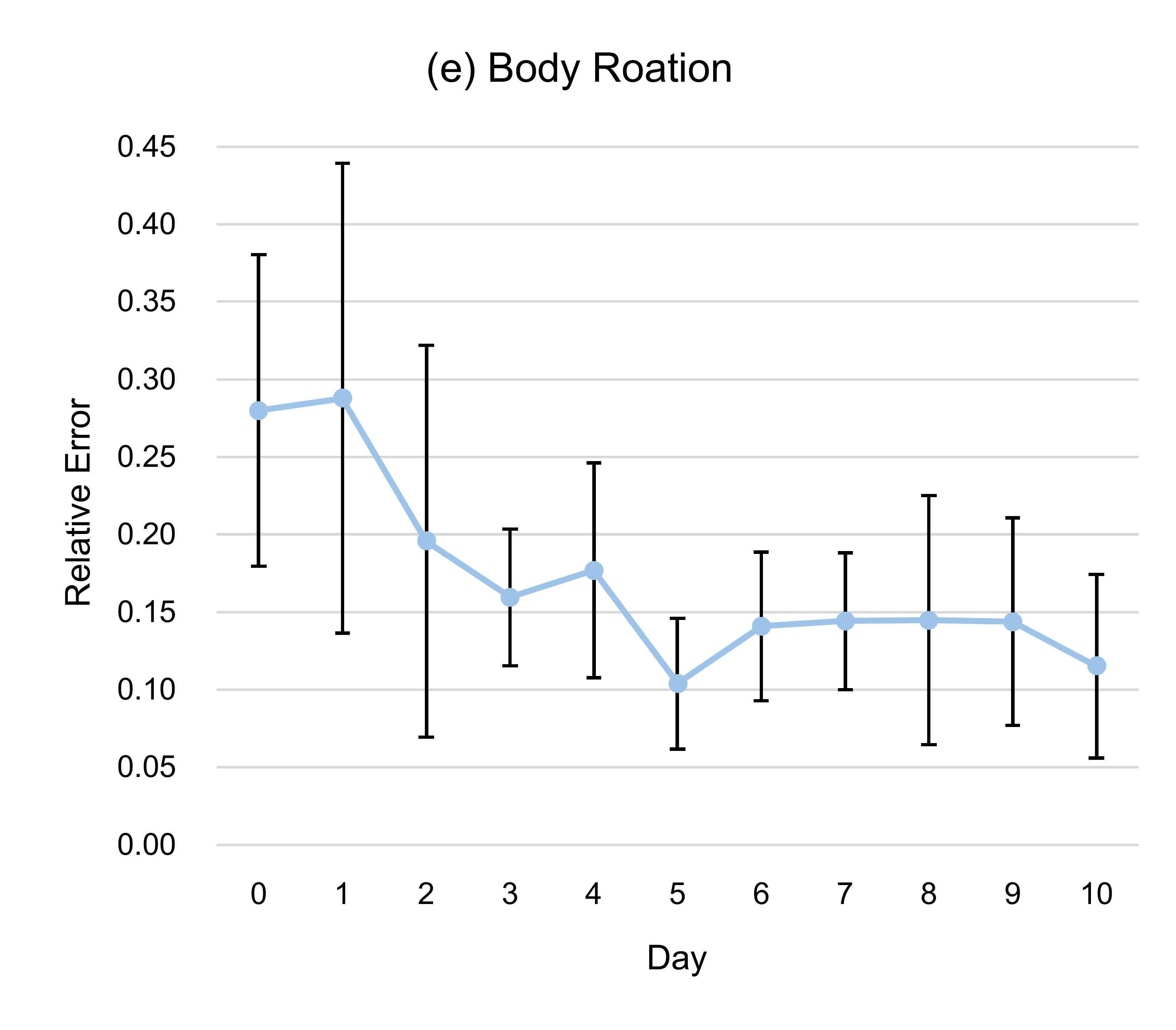}
    \end{minipage}%
    \\
    \caption{The learning curves of each {\itshape Gesture}. }\label{Fig: Each Gesture Curve}
    \Description{The curves of each Gesture type's estimated marginal means.}
    \label{fig:learning curve}
\end{figure*}

Figure \ref{fig:learning curve} indicates the learning curves of each gesture, where \textit{Day 0} indicates the initial pre-study assessment result and the error bars indicate the standard deviations.

\section{Radar of participants' relative errors}

\begin{figure*} [h]
    \centering
    \begin{minipage}[t]{0.28\linewidth}
        \centering
        \includegraphics[width=\linewidth]{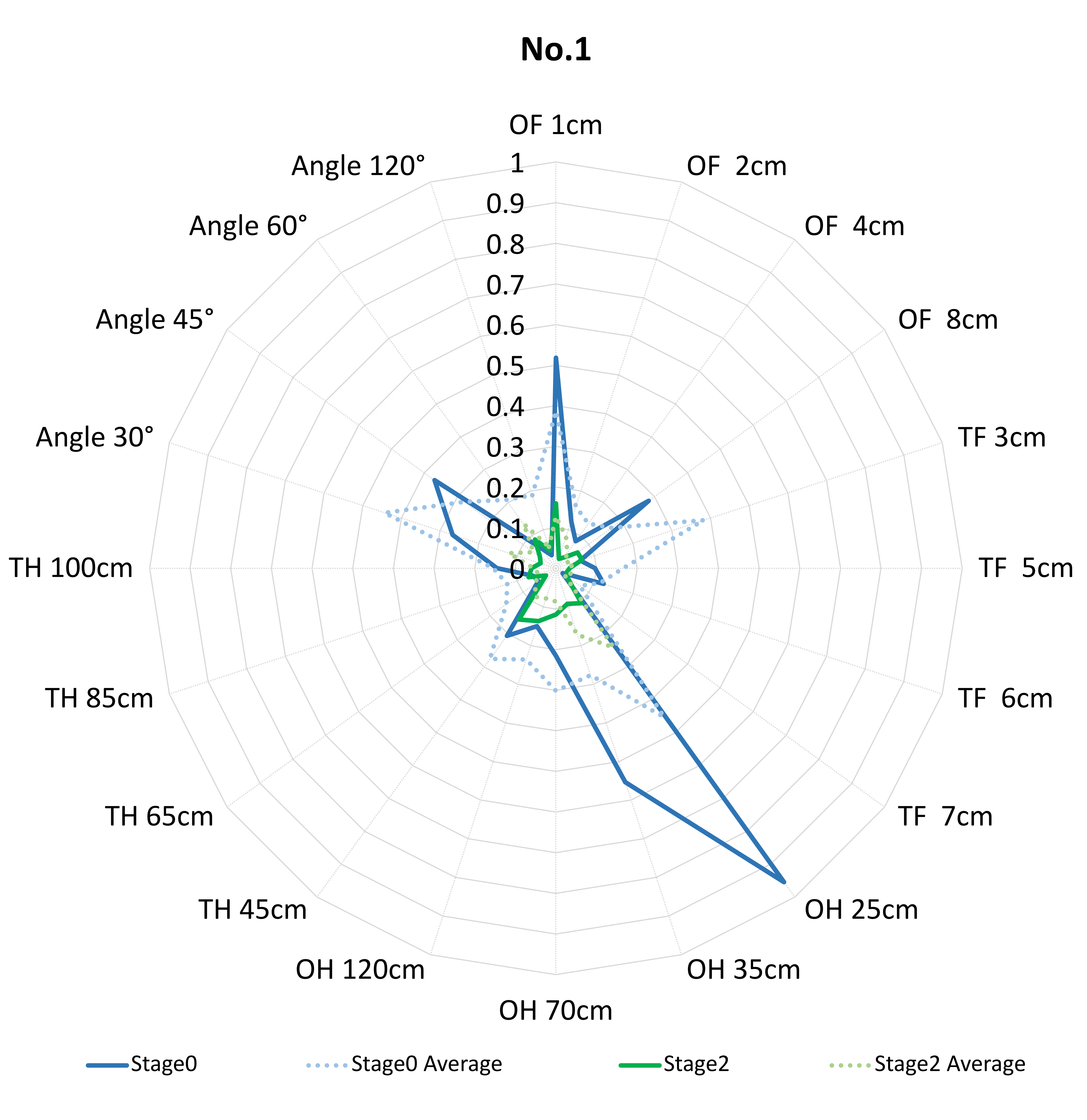}
    \end{minipage}%
    \begin{minipage}[t]{0.28\linewidth}
        \centering
        \includegraphics[width=\linewidth]{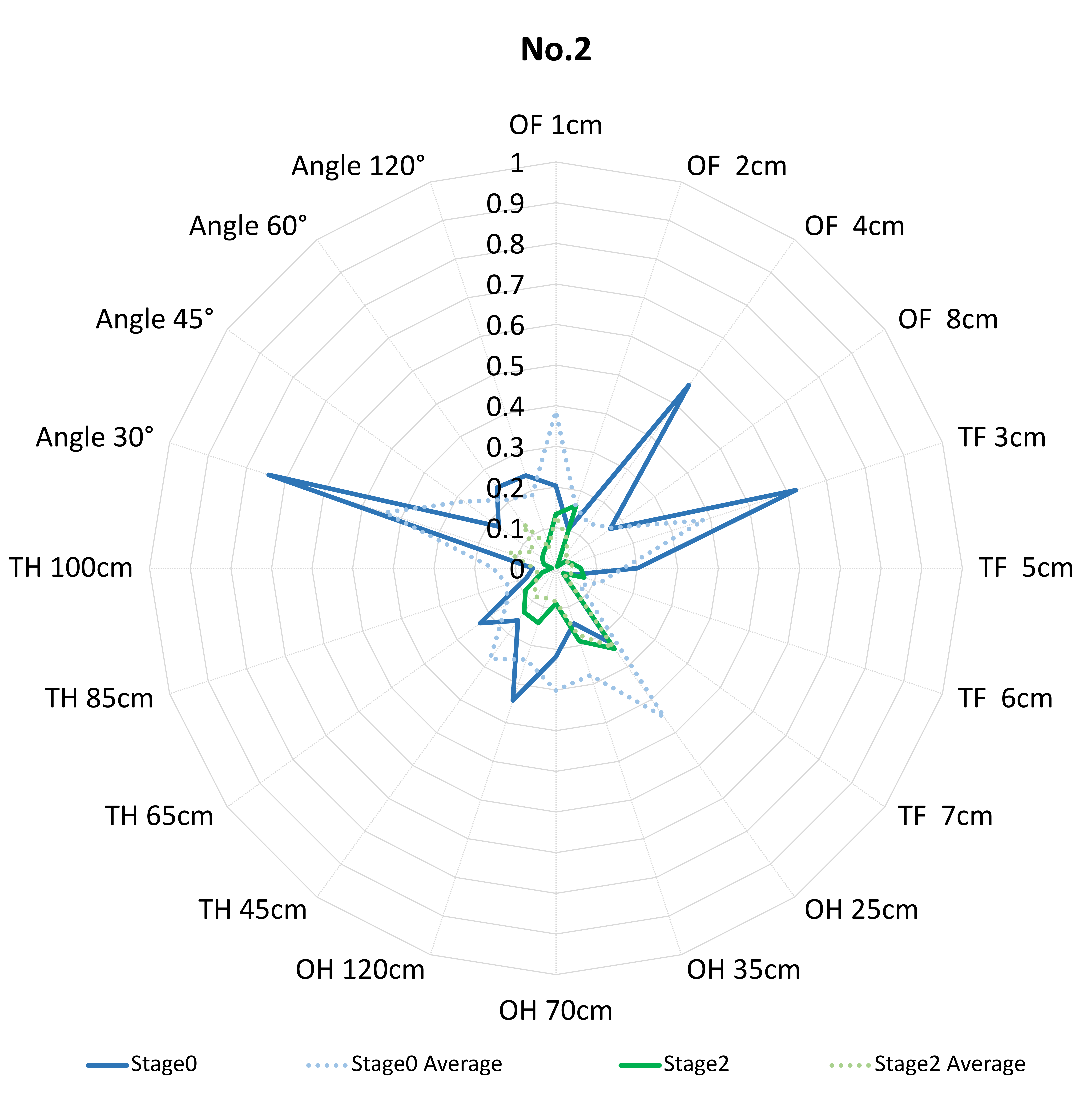}
    \end{minipage}%
    \begin{minipage}[t]{0.28\linewidth}
        \centering
        \includegraphics[width=\linewidth]{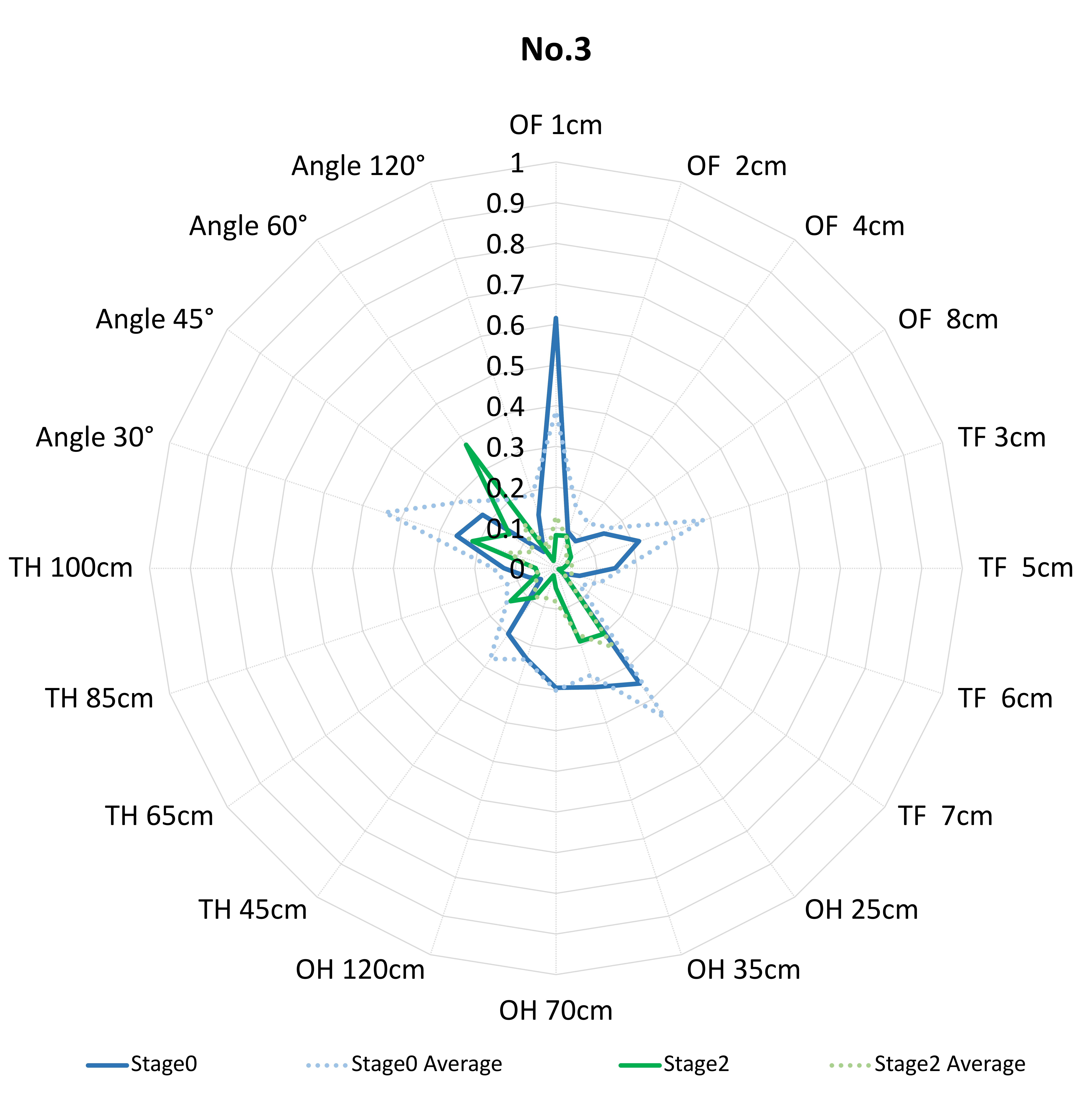}
    \end{minipage}%
    \\
    \begin{minipage}[t]{0.29\linewidth}
        \centering
        \includegraphics[width=\linewidth]{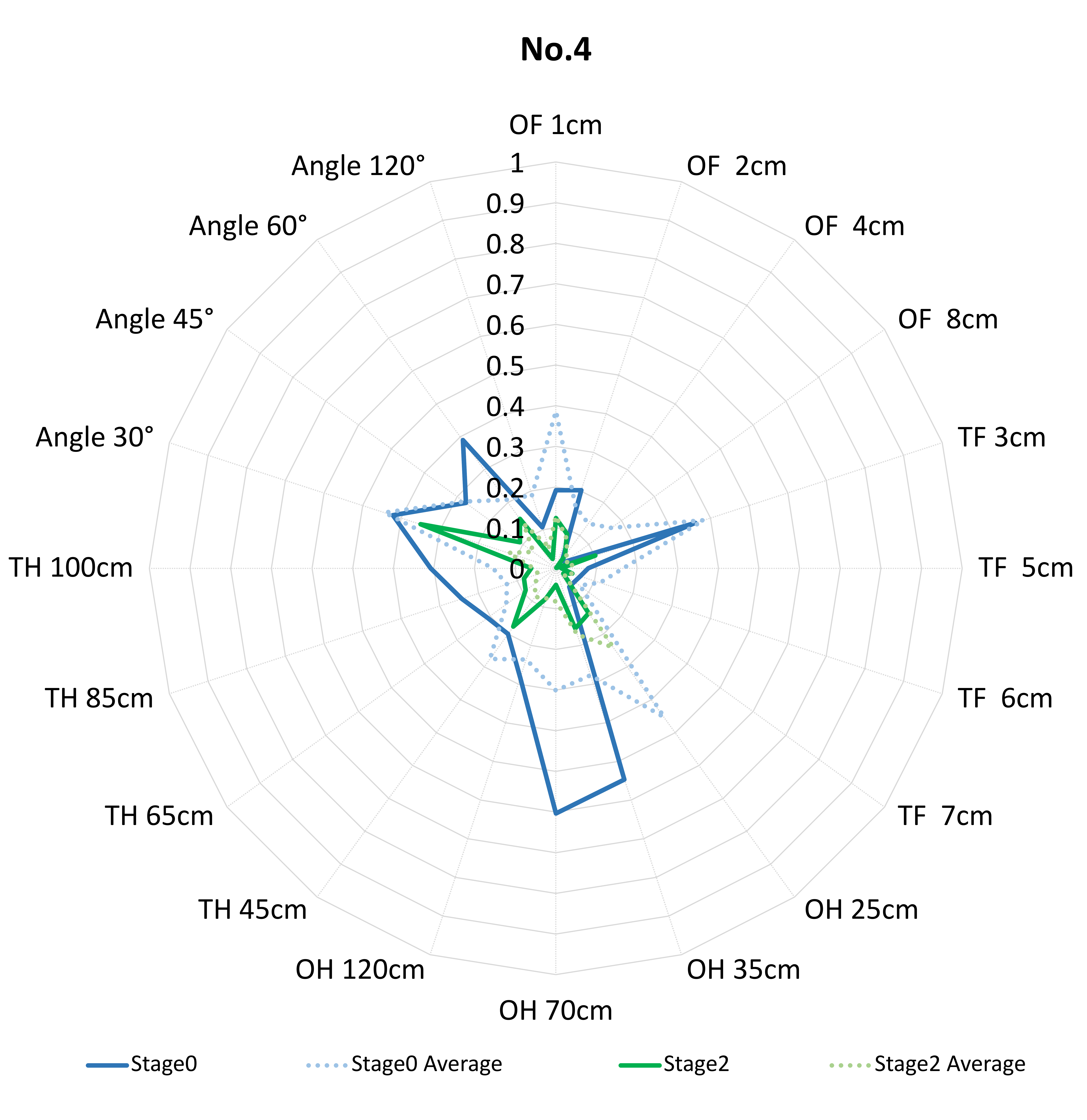}
    \end{minipage}%
    \begin{minipage}[t]{0.29\linewidth}
        \centering
        \includegraphics[width=\linewidth]{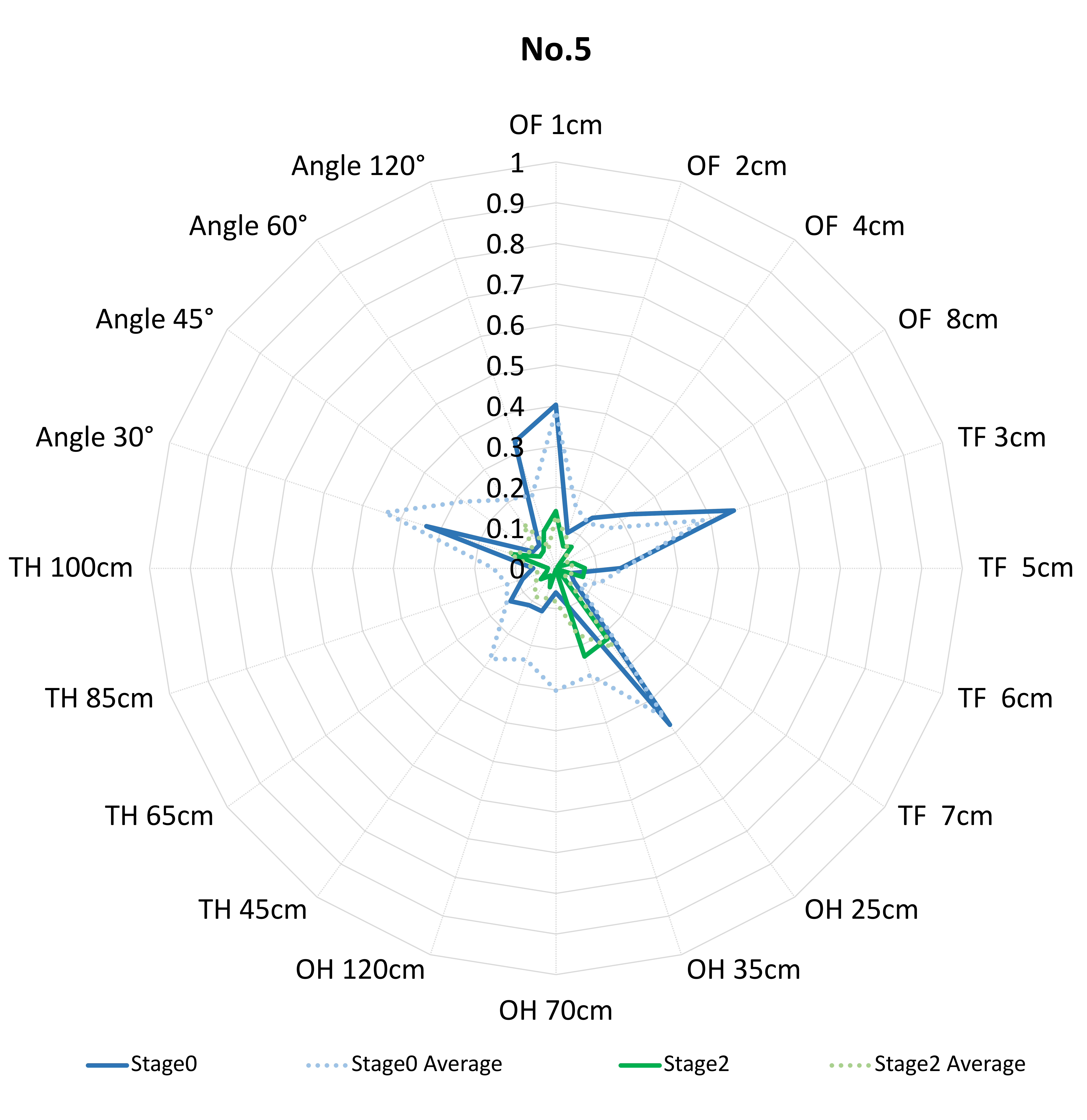}
    \end{minipage}%
    \begin{minipage}[t]{0.29\linewidth}
        \centering
        \includegraphics[width=\linewidth]{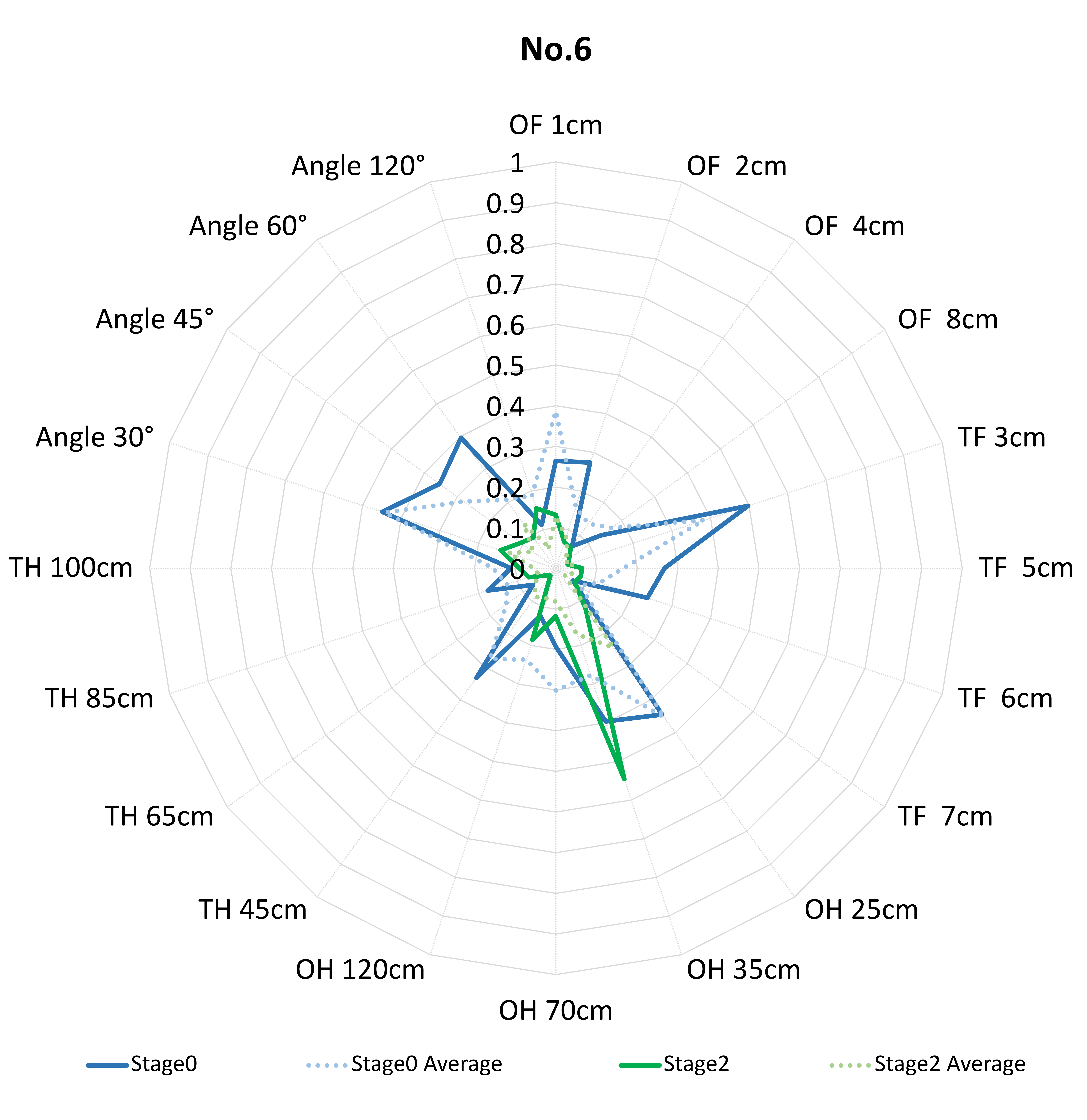}
    \end{minipage}%
    \\
    \begin{minipage}[t]{0.29\linewidth}
        \centering
        \includegraphics[width=\linewidth]{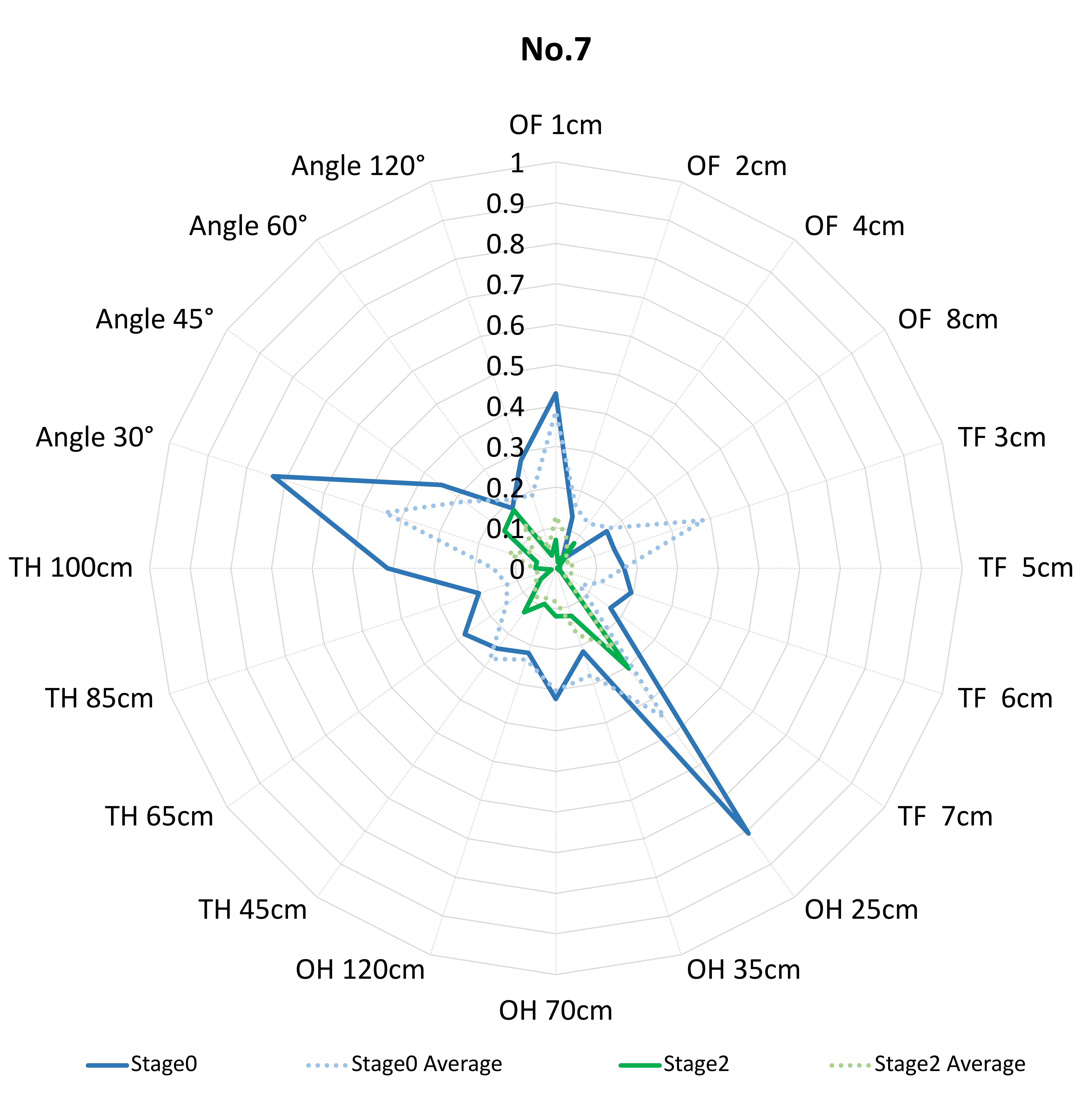}
    \end{minipage}%
    \begin{minipage}[t]{0.29\linewidth}
        \centering
        \includegraphics[width=\linewidth]{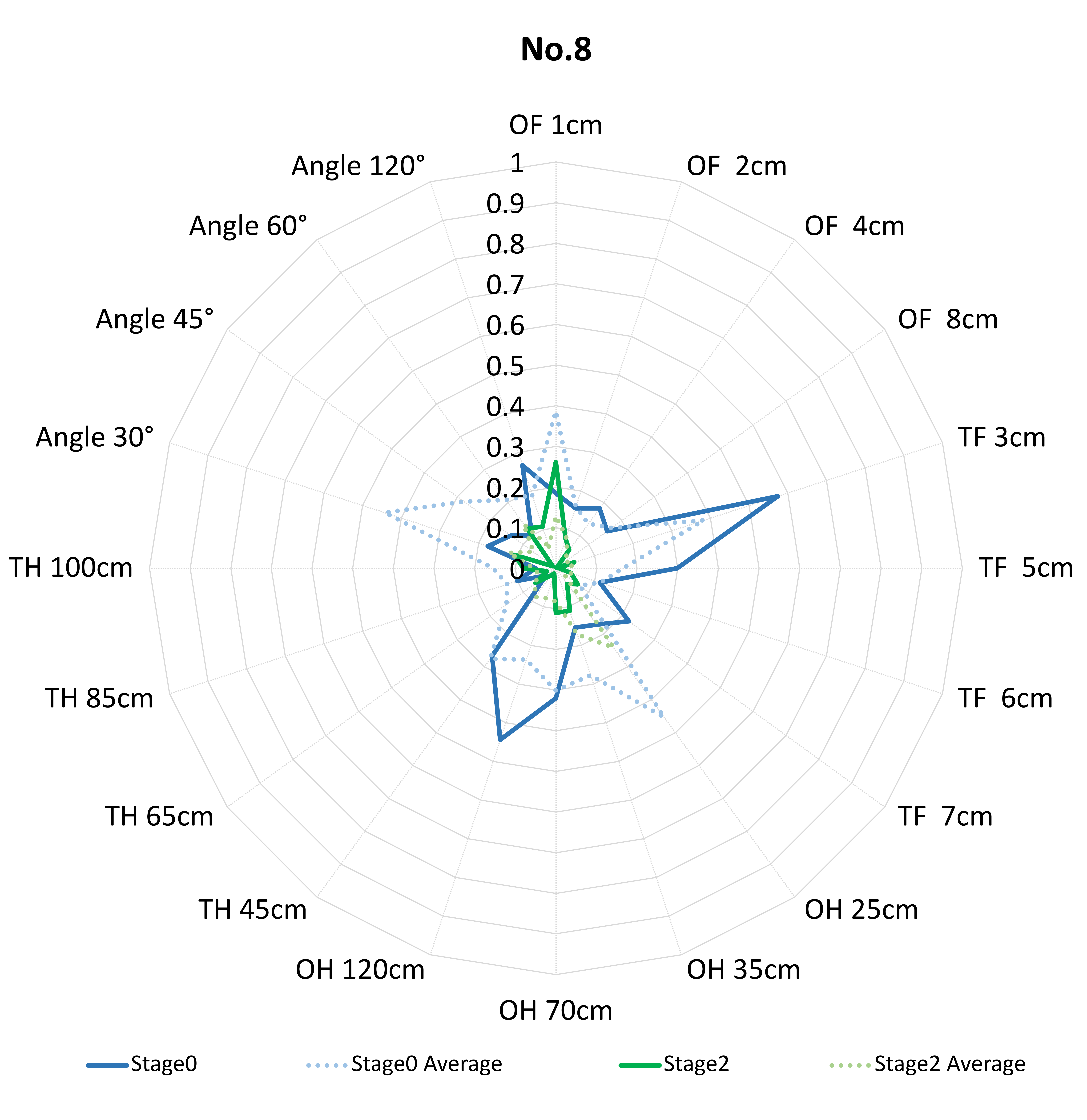}
    \end{minipage}%
    \begin{minipage}[t]{0.29\linewidth}
        \centering
        \includegraphics[width=\linewidth]{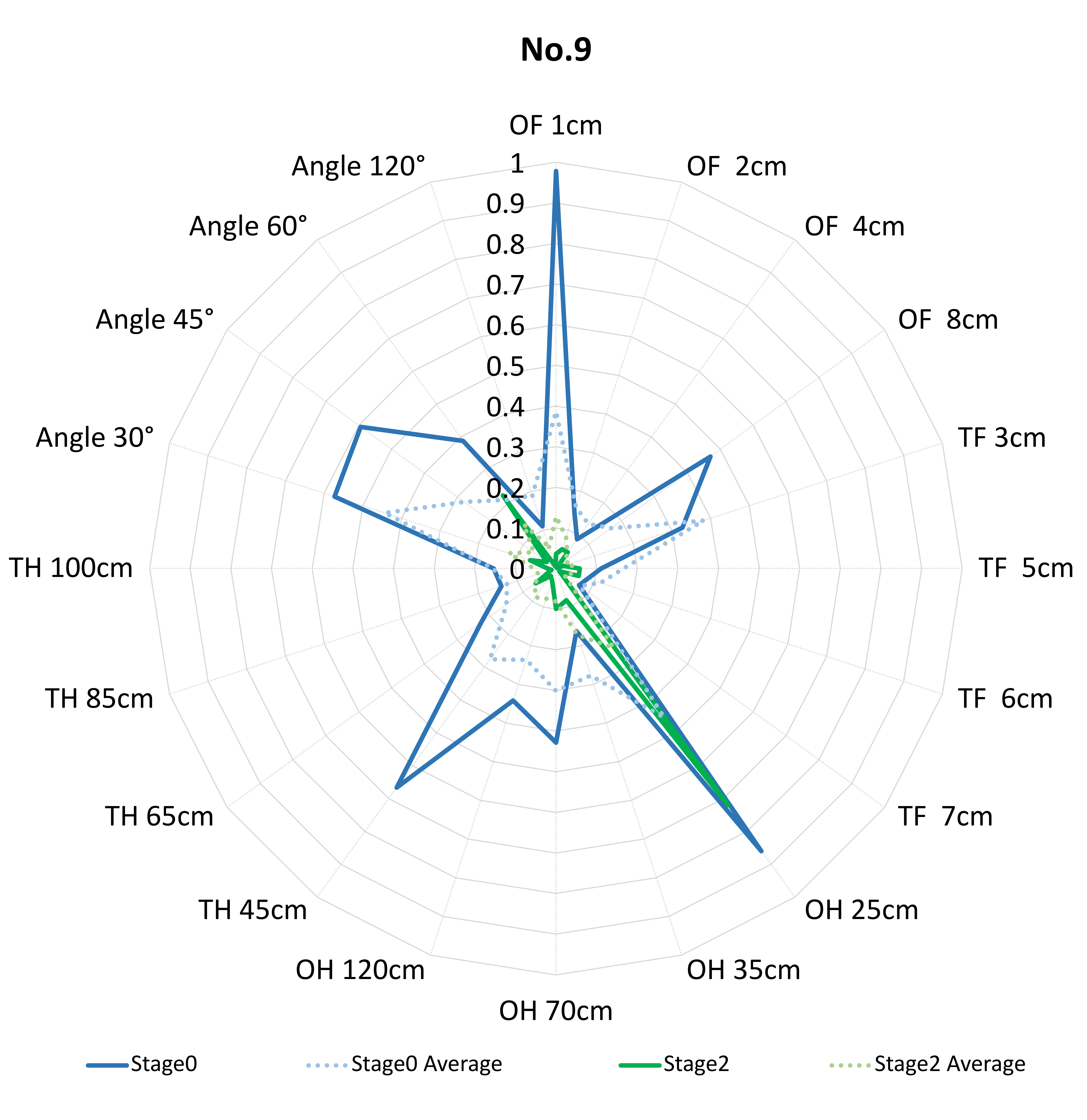}
    \end{minipage}%
    \\
    \begin{minipage}[t]{0.29\linewidth}
        \centering
        \includegraphics[width=\linewidth]{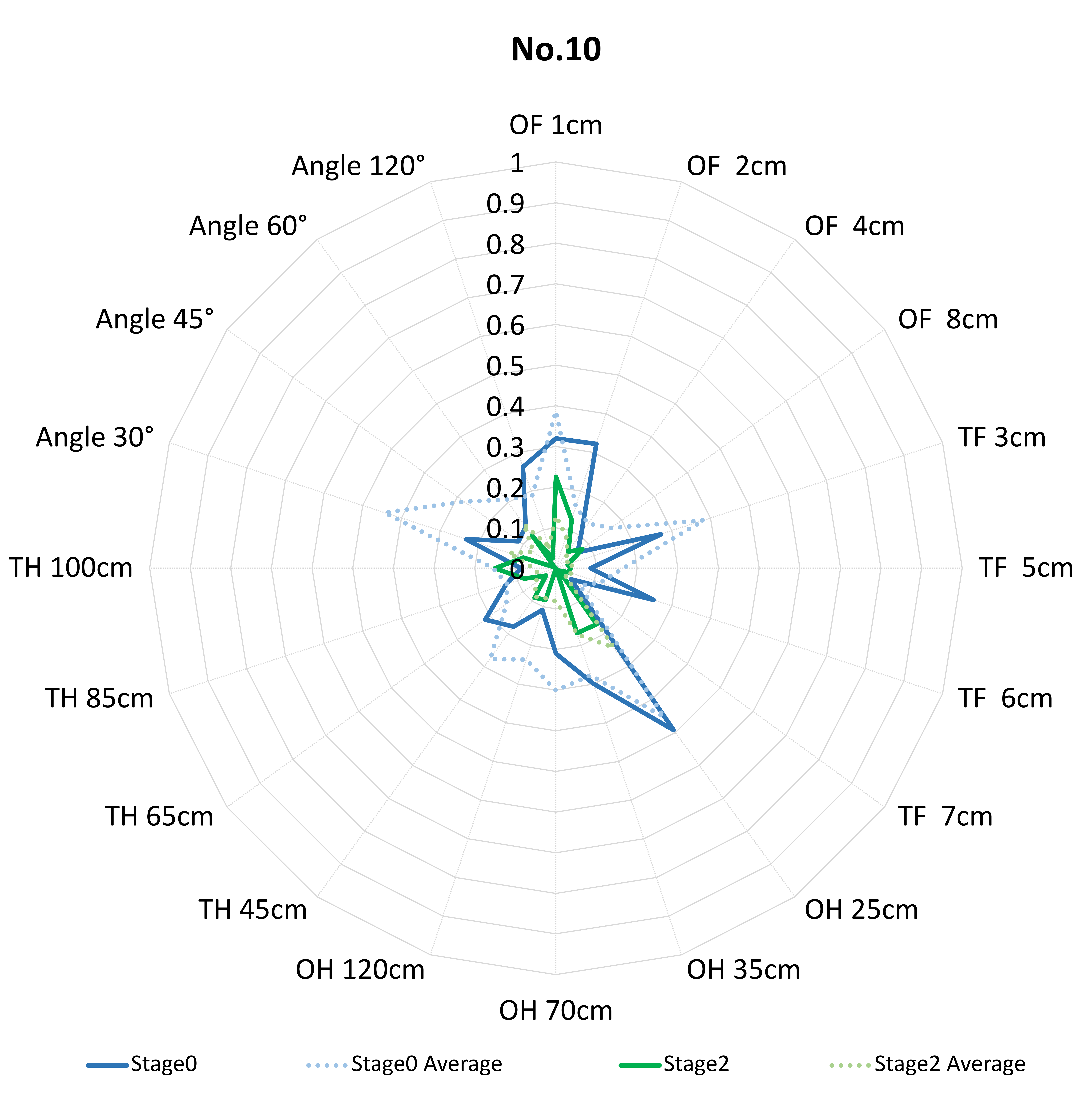}
    \end{minipage}%
    \begin{minipage}[t]{0.29\linewidth}
        \centering
        \includegraphics[width=\linewidth]{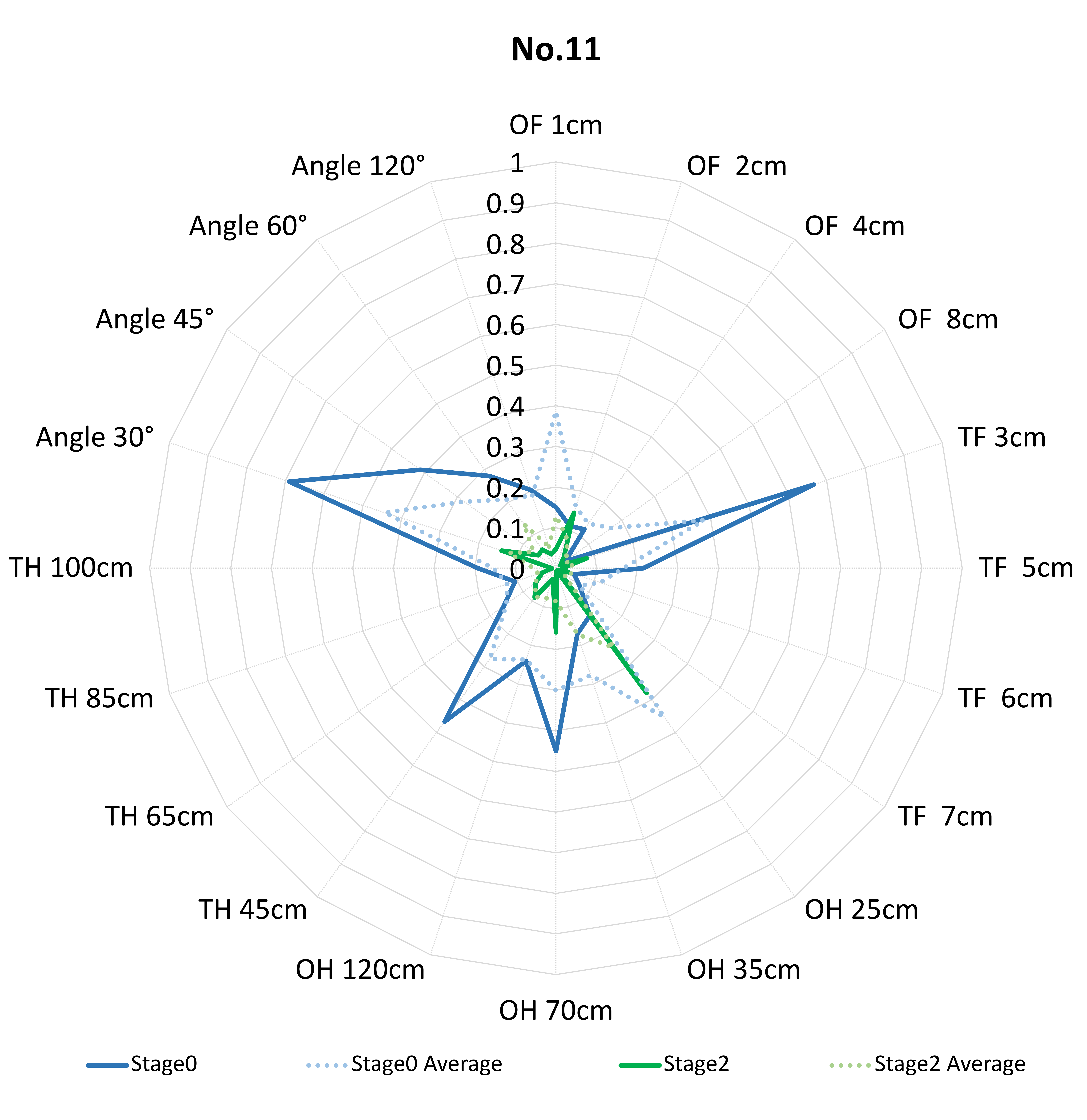}
    \end{minipage}%
    \begin{minipage}[t]{0.29\linewidth}
        \centering
    \end{minipage}%
    \\
    \caption{Radar figures of 11 participants' relative errors on all the tasks.}\label{Fig10: All Radar}
    \Description{Radar plots of 11 participants' relative errors on all the tasks.}
    \label{fig: rader}
\end{figure*}

The radar figures of 11 participants' relative errors on all tasks are shown in Figure \ref{fig: rader}. 




\end{document}